\documentclass[acmsmall]{acmart}

\AtBeginDocument{%
  \providecommand\BibTeX{{%
    \normalfont B\kern-0.5em{\scshape i\kern-0.25em b}\kern-0.8em\TeX}}}

\setcopyright{none}

\usepackage[all,cmtip]{xy}
\usepackage[linesnumbered,ruled,vlined]{algorithm2e} 
\usepackage{enumitem}
\usepackage{multirow}
\usepackage[yyyymmdd,hhmmss]{datetime}
\usepackage{tcolorbox}
\begin{document}


\title{A Categorical Unification for Multi-Model Data: 
\\ Part I Categorical Model and Normal Forms}

\author{Jiaheng Lu}
\affiliation{%
  \institution{University of Helsinki}
  \country{Finland}
}
\email{Jiaheng.lu@helsinki.fi}


\begin{abstract}

Modern database systems face a significant challenge in effectively handling the Variety of data. The primary objective of this paper is to establish a unified data model and theoretical framework for multi-model data management. To achieve this, we present a categorical framework to unify three types of structured or semi-structured data: relation, XML, and graph-structured data. Utilizing the language of category theory, our framework offers a sound formal abstraction for representing these diverse data types. We extend the Entity-Relationship (ER) diagram with enriched semantic constraints, incorporating categorical ingredients such as pullback, pushout and limit. Furthermore, we develop a categorical normal form theory which is applied to category data to reduce redundancy and facilitate data maintenance. Those normal forms are applicable to relation, XML and graph data simultaneously, thereby eliminating the need for ad-hoc, model-specific definitions as found in separated normal form theories before. Finally, we discuss the connections between this new normal form framework and  Boyce-Codd normal form, fourth normal form, and XML normal form.


\end{abstract}

\maketitle

\section{Introduction}

Managing data variety poses a significant challenge in modern database systems due to the diverse formats and models of data sources \cite{Lu:2019:MDN:3341324.3323214,journals/pvldb/KiehnSGPWWPR22}. Currently, relational data adheres to the relational model, XML and JSON data are typically represented using tree-structured models, and property-graph and RDF data are structured as graphs. A natural yet fundamental question is whether there exists a unified data model to unify those different types of data, which will bring together disparate types of data into a coherent framework and lead to a deeper understanding of the diverse data and potentially have practical applications for developing a new type of database.

In this paper, we propose a unifying framework through the lens of category theory to address three types of structured data: relation, XML, and property graph data. Category theory's foundations were established by mathematicians \texttt{Samuel Eilenberg} and \texttt{Saunders Mac Lane} in the 1940s with the primary goal of categorizing and unifying topological objects and algebraic objects. Although category theory originated as an abstract mathematical theory, its applications have extended to various scientific fields, such as physics \cite{coecke2006introducing,zeng2019quantum}, philosophy \cite{peruzzi2006meaning,gangle2015diagrammatic}  and computer science \cite{pierce1991basic,fiadeiro2005categories,shiebler2021category}.

\begin{figure}\centering\includegraphics[width=0.6\textwidth]{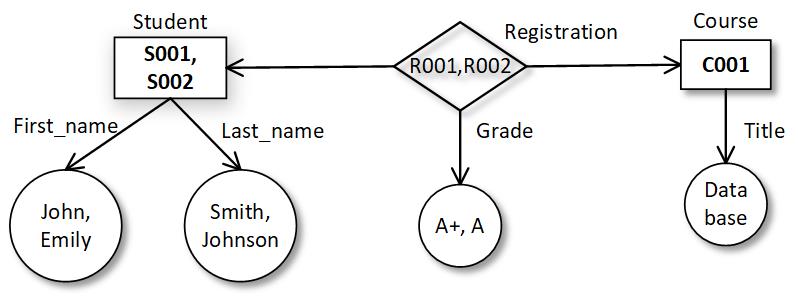}
\caption{This example shows a categorical representation of a toy database, where the object ``\texttt{Student}'' has two attribute objects, ``\texttt{First\_name}'' and ``\texttt{Last\_name}'' A function ``\texttt{First\_name}'' maps the surrogate key $S001$ to ``\texttt{John}'' and $S002$ to ``\texttt{Emily}''. `\texttt{Registration}'' is a relationship object (set) with two projection functions for ``\texttt{Student}'' and ``\texttt{Course}''.} \label{fig:firstexamplecategory}
\end{figure}

A category is composed of \textit{objects} and \textit{morphisms} that establish connections between pairs of objects, known as the domain and codomain of the morphism. Initially, morphisms in category theory were primarily conceptualized as homomorphisms between algebraic structures or continuous maps between topological spaces. However, as categories found applications in diverse domains, researchers began exploring objects and morphisms from domain-specific perspectives. For instance, in physics, objects are defined as states of systems, while morphisms represent transitions from one state to another \cite{coecke2006introducing}. In logic, objects can be represented as propositions, and morphisms serve as a means to demonstrate the inference of one proposition from another \cite{blute2004category}. In this paper, we adopt the category of \textbf{sets}, where each object corresponds to a set, and each morphism represents a function between two sets. Figure \ref{fig:firstexamplecategory} shows an example of a category representation of a toy database. 


In the following, we explore several fundamental concepts in category theory, such as composed morphisms, pullback and limit, providing intuitive examples that demonstrate how these abstract mathematical concepts can be effectively applied to databases.

\noindent \textbf{Composed morphisms}: Let $\mathcal{C}$ be a category, and let $f: A \to B$ and $g: B \to C$ be two morphisms in $\mathcal{C}$. The composed morphism, denoted as $g \circ f$, represents the composite of $f$ and $g$. In our setting, morphisms are defined as functions between sets, so the composition $g \circ f$ is to be thought of as the composed functions.  When there are three composable morphisms, the associativity  $(h \circ g) \circ f$ = $h \circ (g \circ f)$ holds. A \textit{functional dependency} (FD) in databases defines how a given set of attributes determines another set of attributes.  Each morphism in a category naturally corresponds to a function dependency. The composition of morphisms in a category can be understood as the transitivity of functional dependencies. 
 
 

 \smallskip
 
  \noindent \textbf{Pullback}:  The pullback object $P$ represents the ``\textit{most general}'' object that simultaneously satisfies the functions specified by $f$ and $g$. Given a category with objects $A$, $B$, and $C$, and morphisms $f: A \to C$ and $g: B \to C$, the pullback of $f$ and $g$ is an object $P$ along with two morphisms $p_1$: $P \to A$ and $p_2: P \to B$ that satisfy a universal property. Note that pullback objects can be used to define \textit{multivalued dependency} (MVD). To understand it, consider two MVDs: \texttt{Class} $\to\to$ \texttt{Text}, and \texttt{Class} $\to\to$ \texttt{Instructor}. Then the following diagram shows that the relation \texttt{(Class, Text, Instructor)} defines indeed a pullback from the two sub-relations \texttt{(Class, Text)} and \texttt{(Class, Instructor)}. 

\[\xymatrix{(Class, Text, Instructor) \ar[r]^{p_1} \ar[d]_{p_2} & (Class, Instructor) \ar[d]^f \\(Class, Text)  \ar[r]_g & Class}\]

\smallskip

\begin{figure}\centering\[\xymatrix{&*+{SPJ (Limit)} \ar[dl] \ar[d] \ar[dr] \\SP \ar[dr] \ar[d]  & PJ  \ar[dl] \ar[dr] & SJ  \ar[d] \ar[dl]  \\  Product & Supplier & Project}\]
\caption{An example to illustrate join limit.} \label{fig:SPJJoin}
\end{figure}
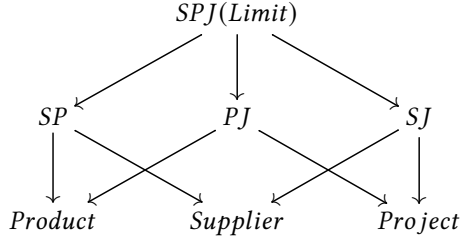

\noindent  \textbf{Limit}:  In category theory, the concept of limit holds significant importance in the analysis and characterization of structures within categories.  While the precise mathematical definition of limits will be provided in subsequent sections, we can illustrate the notion of limit through the scenario involving three tables: \texttt{SP(Supplier, Product)}, \texttt{SJ(Supplier, Project)}, and \texttt{PJ(Project, Product)}. Interestingly, the table \texttt{SPJ (Supplier, Project, Product)} = $SP \bowtie PJ \bowtie SJ$, which joins these three tables, can be described using the concept of a limit. This description can be visually depicted through the corresponding limit diagram in Figure \ref{fig:SPJJoin}.

The true measure of a new framework lies in its ability to yield novel findings. In this paper, the proposed categorical framework makes the following  contributions:

(i)  The framework presented here offers a unique perspective on multi-model data by defining a database as a set category. We illustrate the connection between pullback and the binary join operator (as shown in multivalued dependency), as well as the connection between pushout and the connected components of an undirected graph. Furthermore, considering the duality between pullback (limit) and pushout (colimit) in category theory, an intriguing observation arises: the join operator in relational databases and the computation of connected components in graph databases manifest ``\textit{duality}''  when analyzed through the lens of category theory.




 (ii)  Existing research has proposed a rich set of normal forms for various data models, including relation (e.g. \cite{journals/tods/Fagin77,fagin1979normal,fagin1981normal}), XML (e.g. \cite{DBLP:conf/pods/ArenasL03,journals/tods/VincentLL04}) and object data model (e.g. \cite{journals/tods/TariSS97}). In this paper, our approach leverages the categorical framework to establish a coherent normal form theory across diverse data types, which avoids the necessity for format-specific definitions. This categorical framework presents a fresh perspective for comprehending the theory of normal forms in databases, where the process of devising database normal forms is essentially equivalent to eliminating redundant morphisms and objects that can be derived from others within a category's representation.  When viewed through the lens of category theory, schema normalization becomes ``\textit{representation reduction}'' of a category. In particular, the data schema output from the first reduced representation satisfies the Boyce-Codd Normal Form (BCNF) and XML Normal Form (XML NF) (\cite{DBLP:conf/pods/ArenasL03}), while that from the second reduced representation conforms to  Fourth Normal Form (4NF).

 The contributions and the organization of this paper are summarized as follows:

 \begin{itemize}
     \item We establish the categorical framework for multi-model databases by extending the ER model with a \textbf{thin set} category that includes rich categorical elements such as objects, morphisms, pullbacks, pushouts, and limits, among others. (Section 2).
     \item We show algorithms for the computation of the closure of function dependencies (FDs) and multivalued dependencies (MVDs) on categories  (Section 4).
     \item We build the framework of the normal form theory on categories by designing two levels of reduced representations of categories (Section 5).
     \item We introduce algorithms tailored to facilitate the mapping of a category's representation to a relational data schema, an XML DTD and/or a graph schema (Section 6 and Appendix).
     \item We demonstrate the correlations between the two levels of reduced representations of a category and BCNF, 4NF, and XML NF (Section 7).
     
 \end{itemize}

 This paper is not the first to apply category theory to databases, even in the more specific domain of multi-model databases (e.g. see the previous works \cite{video,journals/jbd/KoupilH22}). For an extensive review of the related literature, please refer to Appendix \ref{sec:relatedwork}. However, note that the contribution of this research lies not only in the development of a categorical model for multi-model databases but also in proposing a cross-model normal form theory for understanding the fundamental principles of multi-model database normal forms.


 %




\section{Thin Set Category}

\nopagebreak

In this section, we describe a formal conceptual data model for multi-model data with a \textbf{thin set category}.

\subsection{Objects and Morphisms in Categories}

\begin{definition}\label{def:category} \cite{MacLane:205493} (Mathematical category) A category $\mathcal{C}$ consists of a collection of objects denoted by $Obj(\mathcal{C})$ and a collection of morphisms denoted by $Hom(\mathcal{C})$.  For each morphism $f \in Hom(\mathcal{C})$ there exists an object $A \in Obj(\mathcal{C})$ that is a domain of $f$ and an object $B \in Obj(\mathcal{C})$ that is a target of $f$. In this case we denote $f \colon A \to B$. We require that all the defined compositions of morphisms are included in $\mathcal{C}$: if $f\colon A \to B \in Hom(\mathcal{C})$ and $g \colon B \to C \in Hom(\mathcal{C})$, then $g \circ f \colon A \to C \in Hom(\mathcal{C})$. We assume that the composition operation is associative and for every object $A \in Obj(\mathcal{C})$ there exists an identity morphism $\text{id}_{A} \colon A \to A$ so that $f \circ \text{id}_{A} = f$ and $\text{id}_{A} \circ f = f$ whenever the composition is defined.
\end{definition}


Intuitively, a category can be viewed as a graph. A category consists of objects and morphisms, where the objects can be thought of as nodes in a graph, and the arrows can be seen as directed edges connecting the nodes. However, a category is a more structured and rich mathematical concept that incorporates additional properties and operations beyond those of a graph. In a category, there is a composition operation defined on the arrows, allowing for the composition of arrows along a path. This composition operation follows certain rules, such as associativity, which are not typically present in a general graph.





A database can be viewed as a category, specifically a \textbf{Set} category. In this context, each object in the category represents a set, while each morphism corresponds to a function between two sets. Both objects and morphisms possess distinct names within a database category (see an example in Figure 1). The composition of morphisms aligns with the composition of functions. Drawing inspiration from the principles of entity-relationship (ER) diagrams, we identify three fundamental types of objects within this category: \textit{entity objects}, \textit{attribute objects}, and \textit{relationship objects}. An entity object can represent a physical object, an event, or a conceptual entity. Visually, an entity object is depicted as a square shape. The elements contained within an entity object consist of a set of surrogate keys, which are sequentially generated integer numbers.  A relationship object captures the connections between various objects. Graphically, a relationship object is symbolized by a diamond shape. Similar to entity objects, the elements within a relationship object are also represented by surrogate keys.  An attribute object is a particular property to describe an entity or a relationship object, which is graphically represented as a circle.  Each attribute is associated with the domain of values. The domain is similar to the basic data types in programming languages, such as string, integer, Boolean, etc.  





Morphisims are defined as functions in sets. A function maps elements from its domain to elements in its codomain. Let $X$ and $Y$ are two sets (objects), then $f$: $X \to Y $ is a function  from the domain $X$ to the codomain $Y$ such that for any element $\forall  x \in X$,  there is certainly one element $y \in Y$, $ y = f(x)$. A morphism is sometimes called an arrow. Thus, we say arrow, function, or morphism interchangeably, and they are equivalent. Given a relationship object $R$ which associates several other objects $X_1$,$X_2$,...,$X_n$, the morphism between $R$ and $X_i$ is called projection morphism, denoted as $\pi_i(R)=X_i$ 
 (1$\leq$$i$$\leq$$n$).   In database theory, a functional dependency (FD) is a constraint that specifies the relationship between sets of values.  In the context of category theory, each morphism, denoted as $X \to Y$, naturally corresponds to a functional dependency. This is due to the fact that there is a unique element in the set $Y$ for every element in the set $X$. The transitivity of functional dependencies can be understood as the composition of arrows in a category.

Similar to a schema for a database table, a category schema provides a descriptive representation of how data within the category is organized and structured. It encompasses various details, such as the category's name, the domains of elements within each object (set) and any constraints (e.g., pullback or pushout to be elaborated upon later) that are applicable to the category.  A schema does not store the actual data within the category but rather defines the overall shape and format of the data. Note that a categorical schema offers a unified view of multi-model data. In this paper, we show algorithms that facilitate the mapping of a categorical schema to different individual data schemas, such as relational schema, XML DTD, and property graph schema.

\subsection{Thin Category}


In this paper, the category under discussion is not just any set category, but rather a thin category, precisely defined as follows:





\begin{definition}(Thin Category) \cite{roman2017introduction} Given a pair of objects $X$ and $Y$ in a category $\mathcal{C}$, and any two morphisms $f$, $g$: $X \to Y$, we say that $\mathcal{C}$ is a thin category if and only if the morphisms $f$ and $g$ are equal. 
\end{definition}





In category theory, a commutative diagram is a graphical representation that depicts the relationships between objects and morphisms.  A diagram $\mathcal{D}$ is commutative meaning that different paths through the diagram yield the same result. In other words, if there are multiple ways to get from one object to another by following a sequence of morphisms, all these paths lead to the same result. The following lemma (\cite{roman2017introduction}) connects thin categories with commutative diagrams.



 \begin{lemma}  [\cite{roman2017introduction}] All diagrams in a thin category are commutative.\label{lem:thincommutative} 
 \end{lemma}

 A thin category ensures the equivalence of various paths, thereby guaranteeing commutativity for all diagrams. It is intriguing to note its link to the Universal Relational Assumption (URA) \cite{DBLP:conf/pods/KuckS82,DBLP:conf/pods/SteinM85} in database theory. The URA assumes that each attribute name must be unique, allowing the consolidation of all information from multiple relations within a single relation through join operators. The output of relational data from a thin category adheres to the URA. Consequently, when examined through the lens of category theory, the URA in database theory can be understood as an alternative way to define a thin category.
 
The following definition summarizes the main characteristics of the database category.


\begin{definition}(Database Category) A database category is a special type of category defined as a thin set category. Each object is a set and each morphism is a function. There are three types of objects: \textit{entity objects}, \textit{attribute objects}, and \textit{relationship objects}. All diagrams in this category are \textbf{commutative}. \end{definition}

\subsection{Multivalued Dependencies and Pullback }
\label{sec:mvd}


While a functional dependency, denoted as $X \to Y$, can be represented as a morphism between the sets $X$ and $Y$ within a category, defining multivalued dependencies (MVDs) requires additional conceptual tools. We employ the concept of \textbf{pullback}, which serves as a universal structure in categories, to define MVDs. 




\begin{definition} (Pullback) Given a category $\mathcal{C}$ with three objects $A$, $B$, and $C$, and two morphisms: $f: A \to B$, $g: B \to C$. The pullback $P$ of $f$ and $g$ consists of morphisms $p_1: P \to A$  and $p_2: P \to B$ such that $f \circ p_1$=$g \circ p_2$  and $P$ has the universal property. That is, given any other object $P'$ with   two morphisms $p'_1: P' \to A$  and $p'_2: P' \to B$ such that $f \circ p'_1$=$g \circ p'_2$, there exists a unique morphism $u: P' \to  P$ with $p_1 \circ u = p'_1$ and  $p_2 \circ u = p'_2$.\end{definition}




In the category of sets, the pullback $P$ of functions $f : A \to C$ and $g : B \to C$ always exists and is given by the join of the sets $A$ and $B$ through $C$.
\[ P = A \bowtie_C B = \{ (a,b) \in A \times B | f(a)=g(b)=c, c \in C \} \]


Given a relation $R(X,Y,Z)$ and two multivalued dependencies (MVDs) $ X \to\to Y$ and  $ X \to\to Z$.   The relation $R$ can be constructed by the following pullback diagram:
\[
\xymatrix{
{XYZ} \ar[r]^{\pi_1} \ar[d]_{\pi_2} & XY \ar[d]^{\pi_3} \\
XZ \ar[r]_{\pi_4} & X
}
\]

\[ XYZ = \{ (x,y,z) \in XY \times XZ |  (x,y) \in XY \wedge (x,z) \in XZ \wedge x \in X \} \]

\subsection{Join Dependency and Limit }
\label{sec:joindependency}


 If $U$ is a universal set of attributes, a join dependency over $U$ is an expression of the form $\bowtie[X_1,...,X_n]$, where each of $X_1,...,X_n$ is a subset of $U$ with the union of the $X_i$’s being $U$. A relation $I$ over $U$ satisfy $\bowtie[X_1,...,X_n]$ if $I = ~\bowtie^n_{i=1}(\pi_{X_i}(I))$. A multivalued dependency (MVD) is a special case of a join dependency.  A relation $I$ over $U$ satisfies an MVD: $X \to\to Y$ if $I = ~\bowtie[XY,X(U-Y)]$.  
 



In category theory,  a limit is considered a broader concept, encompassing and generalizing the notion of a pullback. In database field, join dependency serves as a more inclusive concept compared to multivalued dependency. By recognizing these analogies, we can establish meaningful connections between join dependencies and limits, unveiling the parallels between these two concepts.

\begin{figure}\centering
\[\xymatrix{
& & {S'} \ar@{.>}[d]|u \ar[ddll]|{\psi'_1} \ar[ddl]|{\psi'_2} \ar[ddr]|{\cdots} \ar[ddrr]|{\psi'_n} \ar@/^1pc/[dd]|{\psi'_3} \\
 & & {S}  \ar[dll]|{\psi_1} \ar[dl]|{\psi_2} \ar[d]|{\psi_3} \ar[dr]|{\cdots} \ar[drr]|{\psi_n} \\
T_1 \ar[dr] \ar[d]|{\pi_{11}}  & T_2  \ar[dl] \ar[dr]  \ar[d]|{\pi_{22}} & T_3 \ar[d]|{\pi_{33}} \ar[dl] \ar[dr] & ... \ar[d]|{\pi_{ij}} \ar[dl] \ar[dr] & T_n \ar[d]|{\pi_{nn}} \ar[dl] \\  
A_1 & A_2 & A_3 & ... & A_m
}
\]
\caption{This commutative diagram serves to illustrate the concept of a join limit. For the sake of clarity, the composed morphisms from $S$ to $A_1, ..., A_m$ and from $S'$ to $A_1, ..., A_m$ have been omitted in the diagram.  } \label{fig:joinlimitdiagram}
 \end{figure}
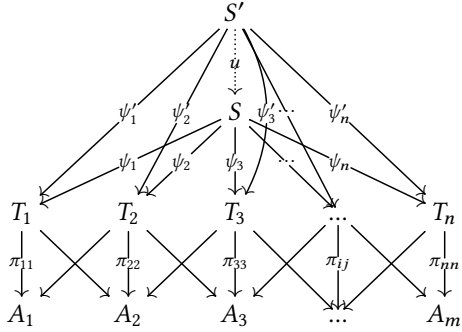



 \begin{definition} (Cone) Let $\mathcal{J}$ and $\mathcal{C}$ be categories. A diagram of $\mathcal{J}$ in $\mathcal{C}$ is a functor $D:$$\mathcal{J}$ $\to$ $\mathcal{C}$. A cone to  $D$ consists of an object $S$ in $\mathcal{C}$ and a family of morphisms in $\mathcal{C}$, $\psi_X: S \to D(X) $ for each object $X$ in $\mathcal{J}$, such that for every morphism $f: X \to Y$, the triangle commutes $D(f) \circ \psi_X = \psi_Y$ in $\mathcal{C}$. $S$ is called the summit of the cone.
\end{definition}

\begin{definition} (Limit) A limit of the diagram $D:$$\mathcal{J}$ $\to$ $\mathcal{C}$ is a cone ($S, \psi_X$) to $D$ such that for every other cone ($S', \psi'_X$) there is a unique morphism $u: S' \to S$ such that $\psi_X \circ u = \psi'_X$ for all $X$ in $\mathcal{J}$. Thus, the limit is the ``\textit{closest}" cone to the diagram $D$. 
\end{definition}






Based on the above general definition of limit, we describe a special type called \textbf{join limit} that is used in this paper to connect join dependency with limit. See the diagram in Figure \ref{fig:joinlimitdiagram}. A join limit is a limit of a diagram indexed by a  category $\mathcal{J}$. Consider a database category $\mathcal{C}$, $T_1,T_2,...,T_n$ are relationship objects, and $A_1,A_2,...,A_m$ are their associate objects. $\pi_{ij}$ is a projection morphisms from $T_i$ to $A_j$ defining a diagram of the shape in $\mathcal{C}$. A cone with the summit $S$ consists of  $m+n$ morphisms $\psi_X$, one for each object $X$ in the indexing category so that all triangles commute. The join limit $(S,\psi_X)$ is the closest cone over the diagram, such that for every other cone ($S', \psi'_X$) there is a unique morphism $u: S' \to S$ such that $\psi_X \circ u = \psi'_X$ for all $X$ in $\mathcal{J}$. In the context of the join limit, all morphisms $\psi_X$ and $\pi_{ij}$ are projection morphisms. The summit $S$ is a join limit, which satisfies $\bowtie[T_1,...,T_n]$. An example of join limit is shown in Introduction section (Figure \ref{fig:SPJJoin}) about three tables on \texttt{Supplier}, \texttt{Product} and \texttt{Project}.

\subsection{Pushout}


In category theory, ``\textbf{duality}'' establishes a relationship between two concepts or operations by interchanging certain elements or properties. One such instance of duality exists between pushout and pullback. While a pullback captures the splitting of a morphism into two through projection, a pushout represents a way to merge two morphisms into a single entity. As pullbacks are associated with join operators, pushouts effectively define the equivalent class within a  database.

\begin{definition}  (Pushout) Given a category $\mathcal{C}$ with three objects $A$, $B$, and $C$, and two morphisms: $f: A \to B$, $g: A \to C$. The pushout $P$ of $f$ and $g$ consists of morphisms $p_1: B \to P$  and $p_2: C \to P$ such that $p_1 \circ f$=$p_2 \circ g$  and $P$ has the universal property. That is, given any other object $P'$ with two morphisms $p'_1: B \to P'$  and $p'_2: C \to P'$ such that $p'_1 \circ f$ = $p'_2 \circ g$, there exists a unique morphism $u: P \to  P'$ with $u \circ p_1 = p'_1$ and  $u \circ p_2 = p'_2$.


\end{definition}

 In the context of set category, the pushout of $f$ and $g$ is the disjoint union of $B$ and $C$, where elements sharing a common preimage (in $A$) are identified, together with the morphisms $p_1, p_2$ from $B$ and $C$, i.e.  $P=(B\cup C)/ \sim $, where $\sim$ is the equivalence relation
such that $f(a) \sim g(a)$ for all $a \in A$ .




The following example illustrates the connection between the pushout and the connected component of an undirected graph.

\begin{example} Given an undirected graph $G$, the \texttt{Edge} table includes two attributes \texttt{Node\_id1} and \texttt{Node\_id2}, which describes edges between any two nodes in $G$.   The pushout object \texttt{Component} computes the connected component in the graph $G$, as illustrated in the following commutative diagram.

\[\xymatrix{Edge \ar[r]^{f} \ar[d]_{g} & Node\_id1 \ar[d]^{p_1} \\Node\_id2 \ar[r]_{p_2} & Component (Pushout) }\]\end{example}


Consider that the pullback object corresponds to the join operator in relational databases, while the pushout object corresponds to the connected component in an undirected graph. As pullback and pushout are dual objects, an intriguing observation arises: \textit{the join operator in relational databases and the computation of connected components in graph databases demonstrate duality when examined through the lens of category theory}.

\section{An overview of categorical normal form}
\label{sec:overview}

\begin{figure}\centering\includegraphics[width=0.9\textwidth]{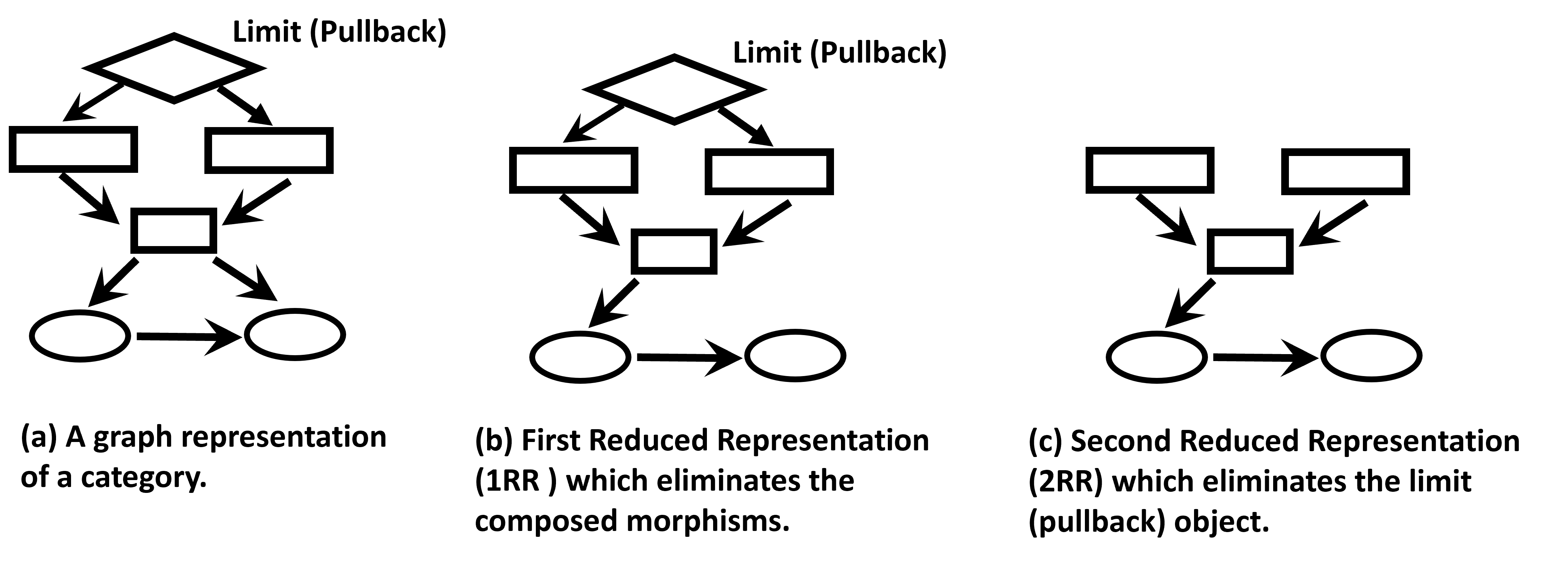}\caption{This figure intuitively illustrates the two reduced representations by removing redundant information. The 1RR removes the composed morphisms and the 2RR further removes the  limit (pullback) object. }\label{fig:reduced}\end{figure}

In this short section, we provide an overview of the categorical normal form theory, which will be described in detail in the subsequent sections.


Figure \ref{fig:reduced} illustrates the main idea behind the normalization process in categories. Given the graph representation of a category, the normalization process involves iteratively removing redundant morphisms and objects from the graph. Specifically, Figure \ref{fig:reduced}(b) shows the first reduced representation, which eliminates morphisms that can be composed from other morphisms, similar to removing transitive functional dependencies in third normal form. Figure \ref{fig:reduced}(c) presents the second reduced representation, which removes the pullback objects, analogous to decomposing a table that can be computed through a lossless binary join in the fourth normal form. We connect the relational normal form  with the reduced representation of categories as follows:

\begin{tcolorbox}[colback=white,colframe=black,width=\linewidth]  

Informally speaking, through the lens of category theory, the essence of normalization for Third Normal Form (3NF) is to eliminate composed morphisms in the graph representation of a category, while the essence of  normalization for Fourth Normal Form (4NF) is to decompose pullback objects.

\end{tcolorbox}

In the following sections, we will establish the corresponding formal results for the above correlations. Specifically, Section 4 develops algorithms to compute the closure of Functional Dependencies (FDs) and Multivalued Dependencies (MVDs) in categories. Section 5 defines the two levels of reduced representations of categories. Section 6 demonstrates how to map a category into relational data. In Section 7, we present the main results, connecting the first and second reduced representations with relational normal form and XML normal form. In summary, the purpose of these sections is to establish a novel, unified normal form theory for databases.

\section{Closure of FDs and MVDs in Categories}


Given the reliance of relational normal form theory on the computation of closures for FDs and MVDs, this section is dedicated to the  presentation of algorithms designed to efficiently compute closures within categories.

\subsection{Closure with FDs}
\label{sec:ClosureFD}

In a relational database, given a set $F$ of FDs, numerous other functional dependencies can be inferred or deduced from the FDs in $F$. In the context of category, if we treat arrows as functional dependencies, then existing arrows can imply new arrows. For example, if $X \to Y$, $Y \to Z$, then $X \to Z$ is an implied (composed) arrow. This is exactly the transitivity rule in Amstrong's axioms \cite{armstrong1974dependency}.  Note that there are three inference rules of functional dependencies in Amstrong's axioms. To compute a closure of a category,  we show how all three Amstrong's axioms can be applied to the category as follows:

\noindent FD 1: If $Y \subseteq X$, then $X \to Y$; 
In this case, $Y$ is an object which is a projection object with a relationship object $X$. We add
 a projection arrow between $X$ and $Y$ in the category.

\noindent FD 2: If $f: X \to Y$, then $g: XZ \to YZ$;  Specifically, given an element $(x,z) \in XZ$, let $y = f(x)$, we define that $(y,z) \in YZ$ and $(y,z)$ is the image of $(x,z)$ under the function $g$. we find an alternative interpretation within the framework of \textbf{monoidal category} for this rule.  A detailed discussion can be found in Appendix \ref{sec:monoidal}. 

\noindent FD 3: If $X \to Y$ and $Y \to Z$, then $X \to Z$. This is the composition rule in categories.



\begin{definition} Given a graph representation $G$ of a category and a set of functional dependencies $F$, any inferred functional dependency $X \to Y$ is considered \textbf{relevant} to $G$ and $F$ if $X$ is either an object in $G$ or appears on the left-hand side (LHS) of a functional dependency in $F$, and $Y$ is an object in $G$.
\end{definition} 

This criterion of relevant functional dependencies ensures that the closure consists solely of the pertinent functional dependencies with respect to $G$.

\begin{definition} 
Given a graph representation $G$ of a category and a set of functional dependencies $F$, the relevant closure representation of $F$ and $G$ denoted as $(G,F)^+$ is a graph where all arrows represent the set of all relevant functional dependencies that can be inferred from $F$ and $G$.

\end{definition}

A schema category $\mathcal{C}$ can be represented as a directed graph, which consists of vertices and arrows, where the vertices correspond to the objects and the arrows indicate the function dependency between these objects. Given a graph representation $G$ of a category and a set of functional dependencies $F$, Algorithm \ref{alg:closureFD} outlines the key steps to compute the closure representation $(G,F)^+$, where each relevant functional dependency is represented as an arrow.


In Line 1 of Algorithm \ref{alg:closureFD}, the graph $G$ is transformed into a set of functional dependencies. Any arrow from $X$ to $Y$ in $G$ is converted to the corresponding functional dependency $X \to Y$. In addition, for each relationship object $X$ in $G$, create two new functional dependencies: $X \to A_1, \ldots, A_n$ and $A_1, \ldots, A_n \to X$, where $A_1, \ldots, A_n$ are projection objects of $X$.   As a result, except for the functional dependencies associated with relationship objects, all other arrows have a single attribute on LHS.  In Line 2, the set $D$ contains all functional dependencies from the category and $F$. Subsequently, for each LHS $X$ of $D$, the closure attributes $X^+$ are computed using Amstong's axioms and the corresponding inferred function dependencies are added in $G$ as arrows (Lines 3-9).






\begin{algorithm}
\caption{Computing the Closure of Categories with FD}
\label{alg:closureFD}
\KwIn{A graph representation $G$ and  a set of functional dependencies $F$ } 
\KwOut{The relevant closure representation  $(G,F)^+$} 
\DontPrintSemicolon

 Convert $G$ to a set of functional dependencies denoted by $FD(G)$;
 
 $D = FD(G) \cup F$;
 
\ForEach{  $X \in LHS(D)$} 
{
    \If{$X \notin G$}
        {
        add $X$ in G; 
        }
    
    Compute the closure attributes of $X$: giving $X^+$;

   \ForEach{  $Y \in X^+$ and $Y \in G$} 
    {
     \If{ there is no arrow from $X$ to $Y$ in $G$}
        {
        add an arrow from $X$ to $Y$ in $G$;
        }
    }

}

\end{algorithm}

\begin{figure}\centering\includegraphics[width=0.7\textwidth]{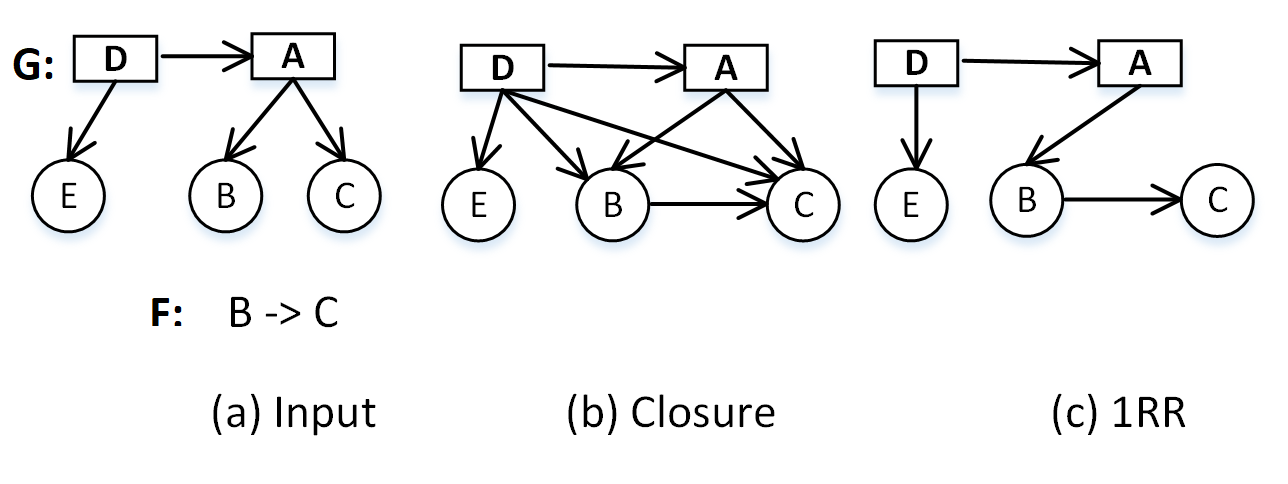}\caption{This example illustrates the compuation of closure and  1RR.}\label{fig:1RRExample}\end{figure}

\begin{example} Figure \ref{fig:1RRExample}(a-b) shows an example to illustrate  Algorithm \ref{alg:closureFD}. Figure \ref{fig:1RRExample}(a) is the input of a category and a FD: $B \to C$.  In Line 1, $FD(G)$ includes four FDs:  $D$$\to$$E$, $D$$\to$$A$, $A$$\to$$B$ and $A$$\to$$C$.  In Line 2, $D = FD(G) \cup \{B \to C\}$. In Line 9, three new arrows are inserted, i.e. $D \to C$, $B \to C$ and $D \to B$ shown in Figure \ref{fig:1RRExample}(b). Note that Figure \ref{fig:1RRExample}(c) will be explained later in Section 
\ref{subsec:1RR}. \end{example}


\subsection{Closure with FDs and MVDs}


In database theory, an MVD $X \to\to Y$ is defined in terms of  ``\textit{a universal set U}''. The reason is that the validity of  $X \to\to Y$ depends on all other attributes associated with this relation $U$.  Unlike previous relational database settings that maintain a fixed $U$, the category setting here allows for different relationship objects having both $X$ and $Y$. As a result, the specification of $U$ becomes an indispensable component of an MVD. To underscore the importance of including the set $U$ in an MVD, we adopt the notation $X \to\to_U Y$ throughout the rest of this paper. This notation serves as a reminder that an MVD is intrinsically tied to the object $U$ within the context of this category.



 While a functional dependency is represented by an arrow, a multivalued dependency is manifested through an \textit{MVD object} in the context of categories. This MVD object works as the universal set $U$ for the corresponding dependency, as defined below:

\begin{definition} (\textbf{MVD objects}) Given a category $\mathcal{C}$, a relationship object $O \in C$ is called an MVD object if there is a multivalued dependency $X \to\to_O Y$ over $C$, such that  $X \cup Y \subset \pi(O) $,  where $\pi(O)$ denote the set of projection objects of $O$. \label{def:MVD}
\end{definition}


Given a set of functional dependencies \( F \) and a set of multivalued dependencies \( M \), the inference rules to compute their closure are well-documented in the literature (see, e.g., \cite{10.5555/551350,10.1145/320613.320614,10.1145/509404.509414}). For the convenience of readers, we have included these inference rules in Appendix \ref{sec:MVDInference}.



\begin{definition} 
Given a graph representation $G$ of a category and a set of functional dependencies $F$ and  a set of multivalued dependencies $M$, the relevant closure representation of  $G, F, M$ denoted as $(G,F,M)^+$ is a graph where all arrows represent the relevant FDs that can be inferred from $F,G$ and $M$, and all MVD objects are identified based on the MVDs that can be inferred from $F,G$ and $M$.

\end{definition}



\begin{algorithm}
\caption{Computing the Closure of Categories with FD and MVD}
\label{alg:closureMVD}
\KwIn{A graph representation $G$, a set of functional dependencies $F$ and a set of multivalued dependencies $M$} 
\KwOut{The relevant closure representation of $(G,F,M)^+$} 
\DontPrintSemicolon

 Convert $G$ to a set of functional dependencies denoted by $FD(G)$;
 
$D$ = $FD(G) \cup F \cup M$;
 
\ForEach{  $X \in LHS(D)$} 
{
        \If{$X \notin G$}
        {
        add $X$ in $G$;
        }

        Compute the FD closure attributes of $X$ based on $D$: giving $X^+$;

        \ForEach{  $Y \in X^+$ and $Y \in G$} 
        {
            \If{$X \to Y $ is not an arrow in $G$}
            {
            add an arrow from $X$ to $Y$ in $G$;
            }
        }

       Identify all MVD objects in $G$ based on the inferred MVDs from $D$;
}

\end{algorithm}

Algorithm \ref{alg:closureMVD} presents the essential steps for calculating the relevant closure representation with functional dependencies $F$ and multivalued dependencies $M$. The algorithm proceeds as follows: In Line 1, the graph $G$ is transformed into a set of FDs. Line 2 initializes a set $D$ to include all the FDs and MVDs. Lines 6 are responsible for computing $X^+$  for each $X \in LHS(D)$ by using $G, F$ and $M$. Subsequently, the new arrows are added, and the MVD objects are identified in Lines 9-10 based on the inferred FDs and MVDs. 



\begin{figure}\centering\includegraphics[width=0.8\textwidth]{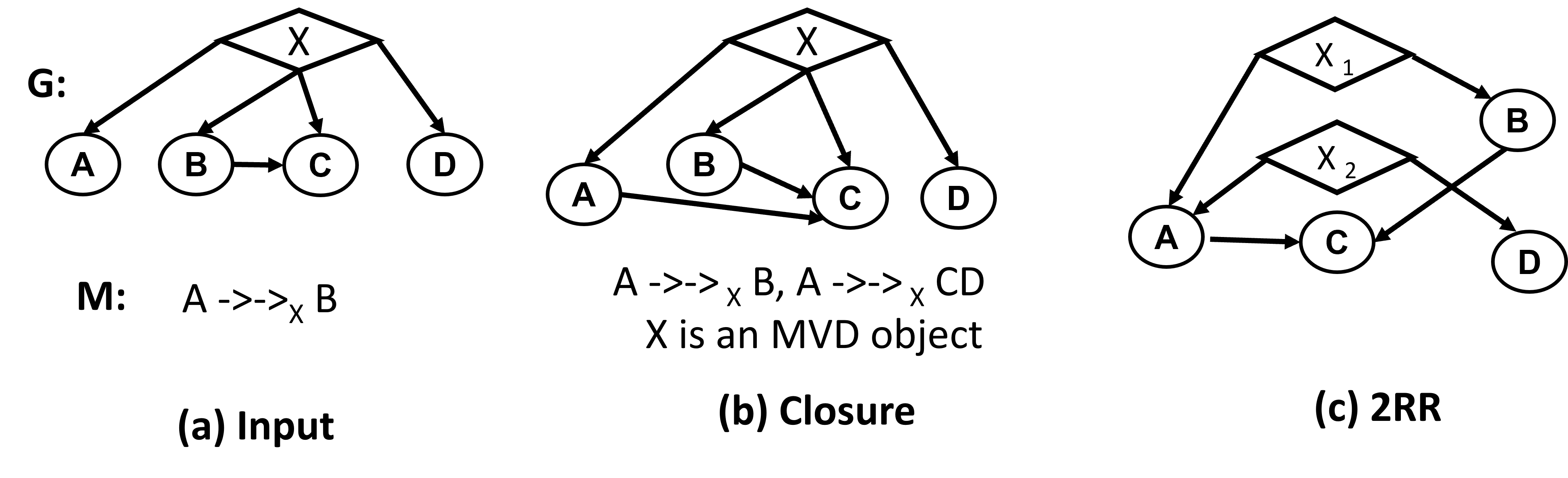}\caption{This example illustrates the closure and 2RR.}\label{fig:2RRExample}\end{figure}

\begin{example}  Figure \ref{fig:2RRExample} (a-b) show an example to illustrate Algorithm \ref{alg:closureMVD}. Fig \ref{fig:2RRExample}(a) is the input with one MVD $A \to\to_X B$. In Line 1, the graph is converted to five FDs. In Line 2, the set $D$ includes five FDs and one MVD.  In Line 9, the arrow $A \to C$ is added due to $A \to\to_X B$, $A \to\to_X CD$ and $B \to C$ (by the inference rule FD-MVD 2). In Line 10, $X$ is identified as an MVD object due to $A \to\to_X B$ and $A \to\to_X CD$, shown in Figure \ref{fig:2RRExample} (b). 
\label{exp:CloureMVDExample}
\end{example}

\section{Representations of Categories}

In this section, we design two levels of reduced representations for categories.

\subsection{First Reduced Representation}
\label{subsec:1RR}

\begin{definition}
Given a set of functional dependencies $F$,  a graph representation $G$ of a category $\mathcal{C}$ is said to cover another graph representation $G'$  if every arrow in $G'$ is also in $(G,F)^+$; that is, if every arrow in $G'$ can be inferred from $G$ and $F$. 
\end{definition}

\begin{definition}
    Given a set of functional dependencies $F$, two graph representations $G$ and $G'$ are equivalent if $G$ covers $G'$ and $G'$ covers $G$. 
\end{definition}

 Roughly speaking, the computation of a First Reduced Representation (1RR) is similar to computing a minimal (canonical) cover of functional dependencies,  wherein an equivalent representation with the minimum number of arrows is sought. Algorithm \ref{alg:1RR} describes the key steps involved in generating the 1RR. The inputs to this algorithm are a graph representation $G$ of a category and a set of functional dependencies $F$ in the canonical form. The output is the 1RR.

\begin{algorithm}
\caption{Computing the First Reduced Representation (1RR)}
\label{alg:1RR}
\KwIn{A graph representation $G$ and  a set of functional dependencies $F$} 
\KwOut{The first reduced representation of $G$} 
\DontPrintSemicolon

 Use Algorithm \ref{alg:closureFD} to compute a closure graph $(G,F)^+$;

\ForEach {arrow  $f: X \to Y$ in $(G,F)^+$} 
{
    \If{($(G,F)^+ - f)$ is equivalent to $(G,F)^+$}
    {
    Remove $f$ in $(G,F)^+$;
    }
}

Return the updated graph $(G,F)^+$;

\end{algorithm}



\vspace{-2mm}
\begin{example}  Recall Figure \ref{fig:1RRExample}, where the subfigure (c) illustrates the removal of three redundant arrows, namely $D \to B$, $D \to C$, and $A \to C$, resulting in the first reduced representation.

\label{exp:1RRExample} \end{example}

\subsection{Second Reduced Representation}



While the first reduced representation condenses the graph representation by eliminating redundant arrows,  the second reduced representation (2RR) takes the optimization process a step further by eliminating (or decomposing) redundant objects, called \textit{derivable relationship objects}.

\begin{definition} (Derivable relationship objects) A  relationship object $O$ in a category $\mathcal{C}$ is derivable if the following conditions are satisfied: (i) $O$ is a limit or an MVD object, and (ii) there is no incoming arrow to $O$; and (iii) for each outgoing arrow $k: O \to X$, where $k$ is not a projection arrow, there exists a projection arrow $f: O \to Y$, and another arrow $g: Y \to X$, such that $k = g \circ f$ in $C$. \label{def:derivable_relation}\end{definition}


A derivable relationship object can be removed without losing any information when the following three conditions are simultaneously satisfied: (i) it is a universal object (limit or pullback); (ii) it has no incoming arrows; and (iii) any outgoing edge can be derived through other arrows. Additionally, if a derivable relationship object is an MVD object (defined in Definition \ref{def:MVD}), it is called a derivable MVD object; otherwise, it is a derivable limit object.

Algorithm \ref{alg:2RR} is designed to generate the 2RR, using a graph representation $G$, a set of functional dependencies (FDs) $F$, and a set of multivalued dependencies (MVDs) $M$ as input. The steps of the algorithm are outlined as follows: The algorithm begins by utilizing Algorithm \ref{alg:closureMVD} to compute a closure category $(G,F,M)^+$ (Line 1). It then focuses on removing derivable MVD objects (Line 2). This is done by iteratively examining MVD objects, denoted as $O$, in the context of $X \to\to_O Y$. When such objects are found, the algorithm decomposes them into two subobjects: $X \cup Y$ and $O - Y$. Subsequently, it eliminates the derivable limit objects. The final steps (Lines 3-5) involve removing the redundant arrows, following a similar approach to the 1RR algorithm.

\begin{algorithm}
\caption{Computing the second reduced representation (2RR)}
\label{alg:2RR}
\KwIn{A graph representation $G$, a set of multivalued dependencies $M$ and  a set of functional dependencies $F$} 
\KwOut{The second reduced representation (2RR) of $G$} 
\DontPrintSemicolon

 Use Algorithm \ref{alg:closureMVD} to compute a closure graph $(G,F,M)^+$;


$G'=$\texttt{Remove\_Objects}($(G,F,M)^+$);

\ForEach {arrow  $f: X \to Y$ in $G'$} 
{
    \If{($G' - f$) is equivalent to $G'$}
    {
    Remove $f$ in $G'$;
    }
}

Return the updated $G'$;




\SetKwFunction{func}{ $\mbox{Remove\_Objects}$}
\SetKwProg{Fn}{Function}{:}{}
\Fn{\func{$G$}}
{


\While{$\exists$ any derivable MVD object $O \in G$ w.r.t $X \to\to_O Y$ }
 {
    decompose $O$ into two objects $O_1= X \cup Y$ and $O_2=O-Y$ and assign the corresponding projection arrows for $O_1$ and $O_2$;

    
  }

  Remove all derivable limit  objects in $G$;

  Return the updated $G$;
}





\end{algorithm}



\begin{example}  This example illustrates the 2RR algorithm. Figure  \ref{fig:2RRExample}(a) is the input. In Line 1, a closure of the category is computed, which has been explained in Example \ref{exp:CloureMVDExample}.  In Line 2,  $X$  is decomposed to $X_1$ and $X_2$ due to MVDs $A \to\to_X B$ and $A \to\to_X CD$.  In Lines 3-5, the redundant arrows are removed. Finally, the 2RR is shown in Fig. \ref{fig:2RRExample}(c), where there is no derivable MVD object. 
\label{exp:2RRExample} \end{example}

\section{Mapping to Relation Data Schema}



\begin{algorithm} \caption{Map  category schema to relational Schema}
\label{alg:map2relationschema}
\KwIn{A reduced graph representation $G(C)$ of  a schema category $\mathcal{C}$ }
\KwOut{Relation schema   $\mathcal{R}$=\{$R_1,R_2,...,R_p$\}}
\DontPrintSemicolon

\ForEach{ unprocessed object $O_i$ in $V(G\mathcal{(C)})$ with outgoing edges } 
{
    Create a table $R_i$ for $O_i$; \tcp*{$i$=1,...,$p$}

    \If{ $O_i$ is a relational or entity object} 
    { 
        $sort(R_i) = \{SK\}$; \tcp*{Surrogate Key}
    }
    \Else{
        $sort(R_i) = \{\lambda(O_i)\}$; \tcp*{$\lambda(O_i)$ is the name of the object $O_i$}
    }

    Add\_neighbours($R_i,O_i$);

    Mark $O_i$ as a processed object;

}

 Clean($\mathcal{R}$);

\SetKwFunction{func}{ $\mbox{Add\_neighbours}$}
\SetKwProg{Fn}{Function}{:}{}
\Fn{\func{$R,O$}}
{
    \ForEach{object $N$ in $outNbr(O)$ }
    {
         

        $sort(R) \leftarrow sort(R) \cup \{\lambda(N)\}$;

        \If{$N$ is a relational or entity object} 
        { 
        Let $\lambda(N)$ be a foreign key in $R$ to reference to $SK$ in $N$; 
        }
        
        \If{ $N \in bin(O)$} 
        { 
            Add\_neighbours($R,N$);   \tcp*{Add the outgoing neighbours of $N$ into $R$}

            Mark $N$ as a processed object;
        }
    }
}

\SetKwFunction{func}{ $\mbox{Clean}$}
\SetKwProg{Fn}{Function}{:}{}
\Fn{\func{$\mathcal{R}$}}
{
    \ForEach{relation $R_i$ in $\mathcal{R}$ }
        {\If{$K$ is the surrogate key without any referential from other relations} 
            { 
              Remove $K$ in $R_i$; 
         } 
    }

    \If{$sort(R_i) \subseteq sort(R_j)$ in $\mathcal{R}$} 
            { 
                Remove $R_i$ in $\mathcal{R}$; 
            } 
        
}
\end{algorithm}

In this section, we present an algorithm that converts a category schema into a relation schema. Due to space constraints, the algorithms that converts a category schema to XML, graph and hybrid schemata are included in Appendixes \ref{sec:XMLDTD} to \ref{sec:hybrid}.




Let us consider a category denoted as $\mathcal{C}$ and its corresponding reduced representation, $G(\mathcal{C})$.  The set of nodes in the representation is referred to as $V(G(\mathcal{C}$)), while the set of edges is denoted as $E(G(\mathcal{C}))$.  For an object $O$ in $V(G(\mathcal{C}$)), the label associated with node $O$ is denoted as $\lambda_G(O)$.  Furthermore, the set of outgoing objects from  $O$ is denoted as $outNbr_{G(C)}(O)$, and the set of bi-directional neighbors is denoted as $bin_G(O)$. To simplify the notation, when the graph $G(\mathcal{C})$ is clear from the context, we may simply use $\lambda(O)$,  $outNbr(O)$, and $bin(O)$ respectively. In addition, a database schema on universe $U$ is a set $R$ of relation schemes \{$R_1,R_2,...,R_p$\},
where  $\bigcup_{i=1}^{p} R_{i} = U$.
The set of attributes of $R_i$ is $sort(R_i)$ and the surrogate key of $R_i$ is denoted by $SK(R_i)$ (if any).

Algorithm \ref{alg:map2relationschema} outlines an approach for the conversion of a reduced representation $G(\mathcal{C})$ into a relational schema.  The algorithm proceeds as follows: For each object $O_i$ having outgoing edges, a corresponding relation $R_i$ is instantiated (Lines 1-8).  The algorithm proceeds to include every outgoing neighbor $N$ of $O_i$ as an attribute within relation $R_i$ (Lines 10-17). If $N$ is an entity or relationship object, the surrogate key of $N$ is treated as a foreign key in $R_i$, referencing the relation generated for $N$ (Lines 13-14). Additionally, if $N$ is a bidirectional neighbor of $O_i$, the algorithm augments $R_i$ with the outgoing neighbors of $N$, since $N$  serves as a key in $R_i$ due to the bijective relationship between $O_i$ and $N$ (Lines 15-17). Finally, the algorithm eliminates subsumed relations and removes any surrogate key that is not referred to by any other relation (Lines 18-23). 


\begin{example} Recall the 1RR graph in Fig \ref{fig:1RRExample}(c). If we run Algorithm \ref{alg:map2relationschema}, the output comprises three relations:  $R_1(A,E)$  (key is $AE$) ($D$ is removed in Line 21), $R_2(A,B)$  (key is $A$) and $R_3(B,C)$  (key is $B$). All those relations satisfy  BCNF. In contrast, if we convert the input graph in Figure \ref{fig:1RRExample}(a) into relations, they include a table $R(A,B,C)$ (with $A$ as the key), which fails to satisfy BCNF (or 3NF) due to $B \to C$.  This example provides an intuitive demonstration of the benefits of the 1RR approach, which ensures that the resulting output schema satisfies BCNF. 

\end{example}

\vspace{-2mm}

\section{Reduced Representation and Normal Forms}

\begin{figure}
\centering
\includegraphics[width=0.7\textwidth]{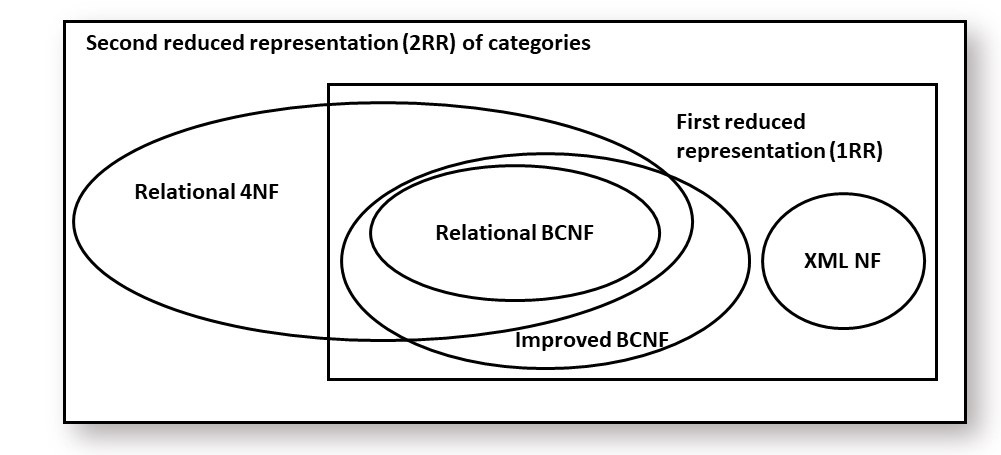}
\caption{Reduced representations of categories and database normal forms} \label{fig:NFScope}
\end{figure}

This section demonstrates that the resulting relational schema from the first and second reduced representations can effectively conform to various levels of normal forms in relational and XML databases.  Figure \ref{fig:NFScope} depicts the structures and relationships among various database normal forms and two reduced representations of categories discussed in this paper.

\begin{theorem} 
Each relational schema $R$ output from the first reduced representation (1RR) in Algorithm  \ref{alg:map2relationschema} is in the Boyce-Codd normal form (BCNF). \label{theo:BCNF}
\end{theorem}



The detailed proofs of theorems can be found in Appendix \ref{sec:proofs}. Further, Ling et al. \cite{journals/tods/LingTK81} showed the inadequacy of BCNF  when applied to multiple relations. They identified that BCNF may contain ``\textit{superfluous}"  attributes. They proposed an enhanced normal form, called \textit{improved} BCNF. The following theorem further builds the connection between 1RR and the improved BCNF.

\begin{theorem} Each relation schema $R$ output from the 1RR in Algorithm  \ref{alg:map2relationschema} is also in the improved  Boyce-Codd normal form as defined in \cite{journals/tods/LingTK81}. \label{theo:improvedBCNF}
\end{theorem}

The output relational schemata of 1RR ensure a certain level of normal form satisfaction. Given its nature as a unified data model, it is reasonable to assume that this categorical framework extends its normalization benefits to corresponding XML and graph data representations. We can establish that the resulting XML DTD, derived from 1RR, satisfies the XML normal form defined by Arenas and Libkin in \cite{journals/tods/ArenasL04}.

\begin{theorem} The DTD schema output D from the first reduced representation is in XML normal form. \label{theo:DTDNormalform}
\end{theorem}

Given that 2RR is a higher level of reduced representation than 1RR, it is intuitively apparent that the data output from 2RR attains a higher level of normalization than that of 1RR. Furthermore, when considering the hierarchy of normal forms,  4NF imposes stricter constraints than BCNF. Thus, it is reasonable to consider that the relational schema resulting from 2RR complies with 4NF, which is described as follows. The proof of the theorem can be also found in Appendix \ref{sec:proofs}.

\begin{theorem} Each relational schema output R from the second reduced representation (2RR) by Algorithm \ref{alg:map2relationschema} is in the fourth normal form. \label{the:4NF}
\end{theorem}


\begin{example}
Recall Fig. \ref{fig:2RRExample}. Running Algorithm \ref{alg:map2relationschema} on Fig. \ref{fig:2RRExample}(a) produces the relation \( R(A, B, C, D) \), which does not satisfy 4NF due to the multivalued dependency \( A \to\to B \) and the functional dependency \( B \to C \). In contrast, computing the 2RR in Fig. \ref{fig:2RRExample}(c) and converting it into relations yields tables \( R_1(A, B) \), \( R_2(A, D) \), \( R_3(A, C) \), and \( R_4(B, C) \), all of which satisfy 4NF. This is a lossless join decomposition and functional dependency preserving.  This example demonstrates the advantage of the 2RR approach in ensuring that the resulting schema follows 4NF. \end{example}


Currently, as far as we know, there is no well-established normal form theory for XML and graph data that directly aligns with the relational 4NF. Consequently, we are unable to prove an analogous theorem for XML or graph 4NF with 2RR. However, the unified framework proposed in this paper enables us to regard the output schema with 2RR as achieving a similar normal form to 4NF for XML and graph data, as illustrated below:

\begin{figure}
\centering
\includegraphics[width=0.7\textwidth]{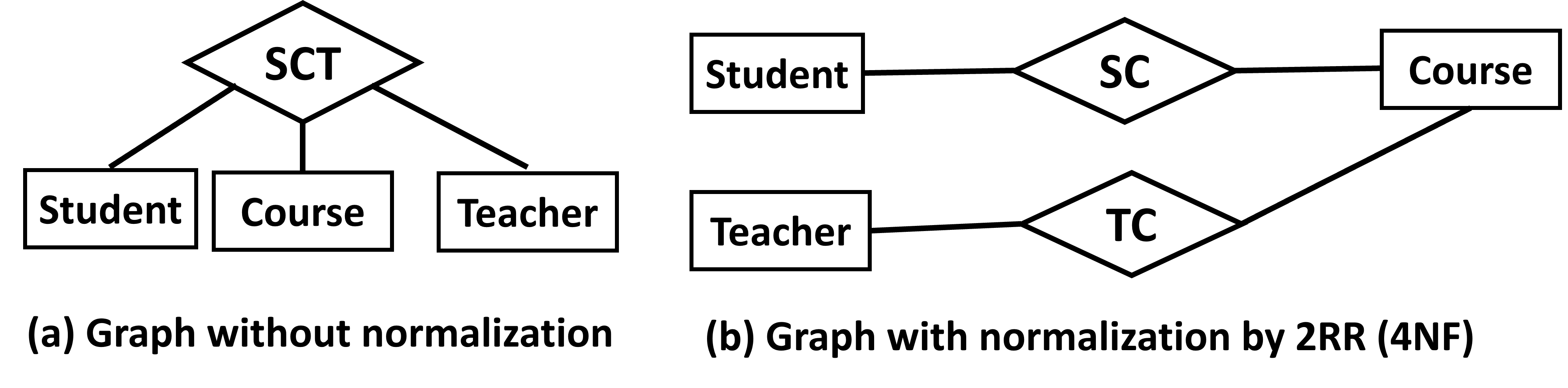}
\caption{An example to illustrate the graph normal form (2RR)} \label{fig:GraphExample2RR}
\end{figure}

\begin{example} This example illustrates a graph normal form with 2RR (corresponding to 4NF). See Figure \ref{fig:GraphExample2RR}. There are two MVDs: \textit{Course} $\to\to$ \textit{Teacher} and \textit{Course} $\to\to$ \textit{Student}. The schema in Figure \ref{fig:GraphExample2RR}(a) contains an $SCT$ node to interconnect three nodes, resulting in duplicated information due to the presence of the two MVDs. In contrast, Figure \ref{fig:GraphExample2RR} (b) is a normalized graph schema based on 2RR, wherein the $SCT$ object is decomposed into two distinct relationship objects ($SC$ and $TC$ nodes). This decomposition effectively eliminates redundancy in the schema. Note that $SC$ and $TC$ node can be removed if the graph schema  supports many-to-many relationships.
\end{example}

\section{Related work}
\label{sec:relatedwork}


The application of category theory to databases has a rich history, with its earliest instances dating back to the 1990s. Notable papers from this period include those by Tuijn and Gyssens (1992,1996) \cite{conf/icdt/TuijnG92,journals/tcs/TuijnG96}, Baclawski et al. (1994) \cite{journals/mscs/BaclawskiSW94}, Libkin (1995)\cite{DBLP:conf/icdt/Libkin95},  and Hofstede et al. (1996) \cite{journals/cj/HofstedeLF96}, etc. While the majority of previous papers focus on the application of category theory to relational databases, there are also works that apply category theory to object databases and graph databases. The existing literature in this field can be broadly categorized into three main areas: \textbf{category theory for conceptual modeling,  for query languages, and for data transformation}. To provide an overview of the related works, Table \ref{tab:relatedwork} presents a taxonomy of these contributions.


\begin{table*}
\caption{Applying category theory on databases}
\label{tab:relatedwork}
\begin{tabular}{ cccc } 
  \toprule
Databases & Conceptual Models & Data Transformation & Query Language  \\ [0.5ex] 
  \midrule
 \multirow{2}{*}{Relational DB} & \cite{journals/ita/LippeH96,143868,piessens1995categorical,conf/amast/JohnsonD93} & \cite{islam1994categorical,DBLP:journals/corr/abs-1009-1166} & \cite{journals/pacmpl/GibbonsHHW18,journals/mscs/BaclawskiSW94} \\
  & \cite{johnson2002entity,journals/cj/HofstedeLF96} & \cite{DBLP:journals/corr/abs-1903-10579,conf/dbpl/SpivakW15} & \cite{johnson2001view,10.5555/2628001,DBLP:conf/icdt/Libkin95,conf/icdt/TannenBW92} \\
  
 Object DB & \cite{journals/tcs/TuijnG96,conf/icdt/TuijnG92}  & None & \cite{conf/pods/OhoriT94,conf/icdt/TannenBW92}  \\ 
 
 Multi-model DB & \cite{conf/vldb/LiuLGHPW18,thiry2018categories,journals/jbd/KoupilH22}  & \cite{conf/er/HolubovaSL19} &   None  \\ 
 \bottomrule 
\end{tabular}
\end{table*}

In order to apply category theory to conceptual models, previous works define concepts in the (extended) ER model using category theory to provide a general and formal framework. In particular, Lippe and Hofstede (1996) \cite{journals/ita/LippeH96} define total role constraints as epimorphisms, unique constraints as monomorphisms, and generalization as colimits. Johnson and Dampney (1993) \cite{conf/amast/JohnsonD93} demonstrate the application of commutative diagrams in developing novel methodologies for constraint specification and process modeling. More recent works by Spivak  (2014) \cite{10.5555/2628001}  consider database schema as a notion of ``onto-logical logs" (olog), which applies the basic concepts of category theory for knowledge representation. 





In the field of category theory applied to query languages, Moggi (1991) \cite{moggi1991notions} proposes to use monads to organize the semantics of programming constructs. Wadler \cite{wadler1990comprehending} (1990) shows that monads are also useful in organizing syntax, in particular, they explain the
"list-comprehension" syntax of functional programming. Libkin and Wong \cite{journals/jcss/LibkinW97} exploit theoretical foundations for querying databases based on bags, showcasing how database operations on collections align with the categorical notion of a monad. Furthermore, Rosebrugh (1991) \cite{rosebrugh1991relational} presents a categorical terminology for describing relational databases, with a particular focus on studying database dynamics by considering updates as database objects within a suitable category indexed by a topos. Libkin (1995) \cite{DBLP:conf/icdt/Libkin95} employs universality properties, a central concept in category theory, to propose syntax for query languages that incorporate approximations.  Gibbons et al. (2018) \cite{journals/pacmpl/GibbonsHHW18} delve into the role of adjunctions, which are categorical generalizations of Galois connections, in the processing of collections. These papers collectively contribute to the field by leveraging category theory to  deepen our understanding of database query languages.

Regarding previous work on the application of category theory to data transformation,  Spivak (2010) \cite{DBLP:journals/corr/abs-1009-1166} demonstrates how morphisms of database schemata give rise to three distinct data migration functors, which serve as powerful tools for translating instances between different schemata in canonical ways. He employs push-forwards to construct joins of tables. In contrast, our paper propose the utilization of limit objects as a means to model join operations, offering an alternative perspective. Furthermore, Spivak and Wisnesky (2015) \cite{conf/dbpl/SpivakW15}  also introduce an algebraic query language, FQL, which is built upon the aforementioned data migration functors.


In the pursuit of applying category theory to object and graph databases, a series of papers have  introduced novel concepts. In particular, Tuijn and Gyssens (1996) \cite{journals/tcs/TuijnG96} introduce typed graphs, where both scheme and
data are defined entirely in terms of categorical constructs; pattern matching of graphs
is realized by morphisms in a suitable graph category.  Ohori and Tajima  (1994) \cite{conf/pods/OhoriT94} show an operation for lifting record operations to objects and classes, which has a similarity with
certain operations associated with monads in category theory. Expanding on the utilization of category theory in database modeling, Tannen, Buneman, and Wong (1992) \cite{conf/icdt/TannenBW92} show the ideas of category theory can be profitably used to organize semantics and syntax in nested relational and complex-object algebras: a cartesian category with a strong monad on it.


In the context of multi-model databases, previous research has explored the application of category theory to  represent multi-model data. In particular, Thiry et al. (2018) \cite{thiry2018categories} propose a categorical approach for modeling relational, document, and graph-oriented models. They demonstrate how to combine category theory with a functional programming language to offer techniques in the context of Big Data. Koupil and Holubova (2022) \cite{journals/jbd/KoupilH22} introduce a mapping between multi-model data and the categorical representation, enabling mutual transformations between different data models. This mapping provides a tool for seamless integration and interoperability of multi-model databases. Furthermore, Liu et al. (2018) \cite{conf/vldb/LiuLGHPW18} envision the structural aspects of multi-model databases and advocate for category theory as a promising new theoretical foundation. Taking a practical perspective, Uotila et al. (2021) \cite{journals/pvldb/UotilaLGLDP21} present a demo system that utilizes category theory for multi-model query processing, employing the Haskell programming language. This demonstration shows the practical implementation and effectiveness of category theory in facilitating multi-model query operations.

While previous pioneering works have explored the application of category theory to various aspects of databases, this paper makes a unique contribution by identifying a connection between the reduced representation of a category and database normal form theory. By establishing this connection, the paper sheds light on the principles for defining a unified theoretical framework for multi-model normal forms.

    Furthermore, it is important to distinguish between the category data model and the object-oriented (OO) data model (e.g. \cite{10.1145/588011.588026,BEERI1990353}). Although both models can be viewed as directed graphs, their semantic definitions differ significantly. The OO model defines objects and methods, while the category model defines objects and morphisms (functions). Unlike methods, morphisms have more rigorous mathematical definitions. Category theory offers a comprehensive set of precise mathematical terminologies and theories, which can lead to valuable insights for databases.




\section{Future work and Conclusion}
\label{sec:futurework}

The presented framework opens up numerous opportunities for future research:

\begin{itemize}[noitemsep,topsep=0pt]
  \item A key aspect to explore is the design of a multi-model query language based on this framework. While relational algebra and calculus have traditionally catered to relational data, leveraging the unifying potential of a category makes it natural to develop categorical algebra and calculus for implementing multi-model queries. Our second paper, ``\textit{Part II Categorical Calculus and Algebra}'', which has also been submitted to the PODS conference, seeks to address this challenge by proposing a rigorous approach to implementing such queries.

  \item We plan to design a unified query optimization plan that can effectively handle different data models. Given that each data model possesses distinct characteristics, the development of a comprehensive optimization strategy capable of accommodating the complexities of multi-model queries becomes crucial.
  \item In this study, a \textit{thin} category is employed to model databases, which follows the \textit{Universal Relational Assumption (URA)}. Nonetheless, it is important to note that the URA may not hold in numerous practical scenarios \cite{10.1145/588111.588113}. Hence, exploring a general non-thin category presents an intriguing avenue for future investigation as well.
\item This paper examines the relationship between the second reduced representation (2RR) and the fifth normal form (5NF) (or called projection-join normal form PJNF \cite{10.1145/582095.582120}). Since join dependencies can be defined using limits, and 5NF is fundamentally based on join dependencies, it is natural to investigate whether a relational schema derived from 2RR that decomposes the limit objects (in Line 10 in Algorithm \ref{alg:2RR}) can satisfy the requirements of 5NF. However, it is important to note that there are no sets of sound and complete inference rules to reason all join dependencies  \cite{10.1145/322307.322313}. Therefore, the 5NF can be satisfied only if there are feasible algorithms to find the closure of join dependencies. 


\end{itemize}

Beyond the technical results, this paper can be regarded
also as arguing that a straightforward application of category theory to databases may \textbf{not} yield optimal results. The primary challenge lies in effectively integrating the language of category theory with existing database theories. For example, the normal form theory proposed in this paper leverages the inference rules of functional and multivalued dependencies in database theory. These rules extend well beyond the composition morphisms employed in a mathematical category.  By navigating the intersection of category theory and database theory, the purpose of our research is to merge the two domains, leveraging the strengths of each to establish a coherent framework for multi-model data management.

\smallskip

\bibliographystyle{ACM-Reference-Format}
\bibliography{sample-base}

\appendix


\newpage

\noindent \textbf{Organization of the Appendixes}  ~~ 


In Appendixes \ref{sec:XMLDTD} to \ref{sec:hybrid}, we show the algorithms used for mapping categories to XML DTD, graph, and hybrid schemata. In Appendix \ref{sec:monoidal}, we analyze the inference rules of FDs and MVDs.  Appendix \ref{sec:proofs} provides the proofs and explanations for the lemmas and theorems. Furthermore, Appendix \ref{sec:complexity} assesses the computing complexities associated with the various algorithms presented.





\section{Mapping Categories to XML DTD}
\label{sec:XMLDTD}

XML is represented as a finite rooted directed tree, with each node in the tree associated with a tag. We use DTD to describe the schema of XML.


\begin{definition} A DTD is a 5-tuple($L, T, P, R, r$) where 
 \begin{itemize}[noitemsep,topsep=0pt]
  \item $L$ is a ﬁnite set of element types (tags);
  \item $T$ is a ﬁnite set of strings (attributes), starting with the symbols @. There is a special attribute @ID, which serves as a unique identifier (key) for a particular tag;
  \item $P$ is a mapping from $L$ to element type definitions: Given any $a \in L$, $P_a = \#P$ or  $P_a$ is a regular expression over $L$ - \{$r$\}, where $\#P$ denotes $\#PCDATA$;
  \item $R$ assigns  each $e \in L $ a ﬁnite  subset of $T$ (possibly empty); 
   \item $r \in L$ is the root.
\end{itemize}
\end{definition}




Algorithm \ref{alg:map2DTD} presents the mapping process from a categorical schema to a DTD schema. The algorithm proceeds as follows: Initially, $\epsilon$ is designated as the root node, and the algorithm processes and adds objects $O$ as children of $\epsilon$ (Lines 1-3).  A new tag $\lambda(O)$ is created for each object with outgoing edges, and $\lambda(O)^+$ is appended to the regular expression of $\epsilon$ (Lines 5-9), where $\lambda(O)$ denotes the label (name) of the object $O$. Subsequently, for each neighbor object $N$ of $O$, if $N$ has any outgoing edge, an attribute $@\lambda(N)$\_ID is added for $O$ (Lines 13-14), otherwise $\lambda(N)$ is appended to the regular expression rule of $\lambda(O)$ (Lines 16-17).  It is worth noting that the shape of the XML data that conforms to the output DTD is wide and shallow. This representation allows for a clear and concise XML structure, avoiding redundant storage.






\begin{example} Recall the 1RR graph in Figure \ref{fig:1RRExample}(c).   Algorithm \ref{alg:map2DTD} convert it to XML DTD as follows: $L=\{A,B,C,D,E\}$, $T$=\{@ID, @A\_ID, @B\_ID\}, $P=\{\epsilon \to D^+A^+B^+, D \to E,  B \to C$\}, $R(A)$=\{@ID, @B\_ID\}, $R(D)$=\{@ID, @A\_ID\}, $R(B)$=\{@ID\}, and $r = \epsilon$. $B$ has an outgoing edge and is appended into the regular expression following the root $\epsilon$. In contrast, $E$ and $C$ are attribute objects without outgoing edges, and thus are created as the sub-element of $D$ and $B$ respectively.  
\end{example}

\begin{example} Recall the 2RR graph in Figure \ref{fig:2RRExample}(c).
Algorithm \ref{alg:map2DTD} convert it to XML DTD as follows: $L=\{A,B,C$, $D,X_1,X_2\}$, $T=$\{@ID,@A\_ID, @B\_ID\}, $P=\{\epsilon \to A^+B^+X_1^+X_2^+$, $A \to  C$, $B \to C$, $X_2 \to D$\}, $R(X_1)$=\{@ID, @A\_ID, @B\_ID\}, $R(X_2)$=\{@ID,@A\_ID\}, $R(A)=R(B)=\{@ID\}$, and $r = \epsilon$.  
\end{example}

\begin{algorithm} \caption{Map  category schema to XML DTD}\label{alg:map2DTD}\KwIn{A graph representation $\mathcal{G(C)}$ of a schema of a category $\mathcal{C}$ }
    \KwOut{XML DTD $\mathcal{D}$=($L$, T, P, R, r)}
    \DontPrintSemicolon

Let $\epsilon$ be the root $r$ of $D$;

 $T= \{@ID\}$;

 Initialize $L$, $P$, and $R$ to be empty sets;

\ForEach{ unprocessed object $O$ in $V(\mathcal{G(C)})$ with outgoing edges}
{ 
     $L \leftarrow L \cup \{\lambda(O)\}$; \tcp*{$\lambda(O)$ is the label of $O$}

     $P_\epsilon \leftarrow P_\epsilon \cdot (\lambda(O))^+$;
    
     $R(\lambda(O))=\{@ID\}$;
    
    Add\_neighbours($O$);
      

     Mark $O$ as a processed node;
    
}

\SetKwFunction{func}{ $\mbox{Add\_neighbours}$}
\SetKwProg{Fn}{Function}{:}{}
\Fn{\func{$O$}}
{
    \ForEach{unprocessed object $N$ in $outNbr(O)$ }
    {    
     \If{$N$ has any outgoing edge}
     { 
        $T= T \cup \{@\lambda(N)$\_ID\};
        
        $R(O)=R(O) \cup \{@\lambda(N)$\_ID\};
     }
    \Else 
     {  
       $P_O \leftarrow P_O \cdot \lambda(N)$;

        $L \leftarrow L \cup \{\lambda(N)\}$;
      }   
    }
}

\end{algorithm}



\begin{figure*}
\centering
\includegraphics[width=0.9\textwidth]{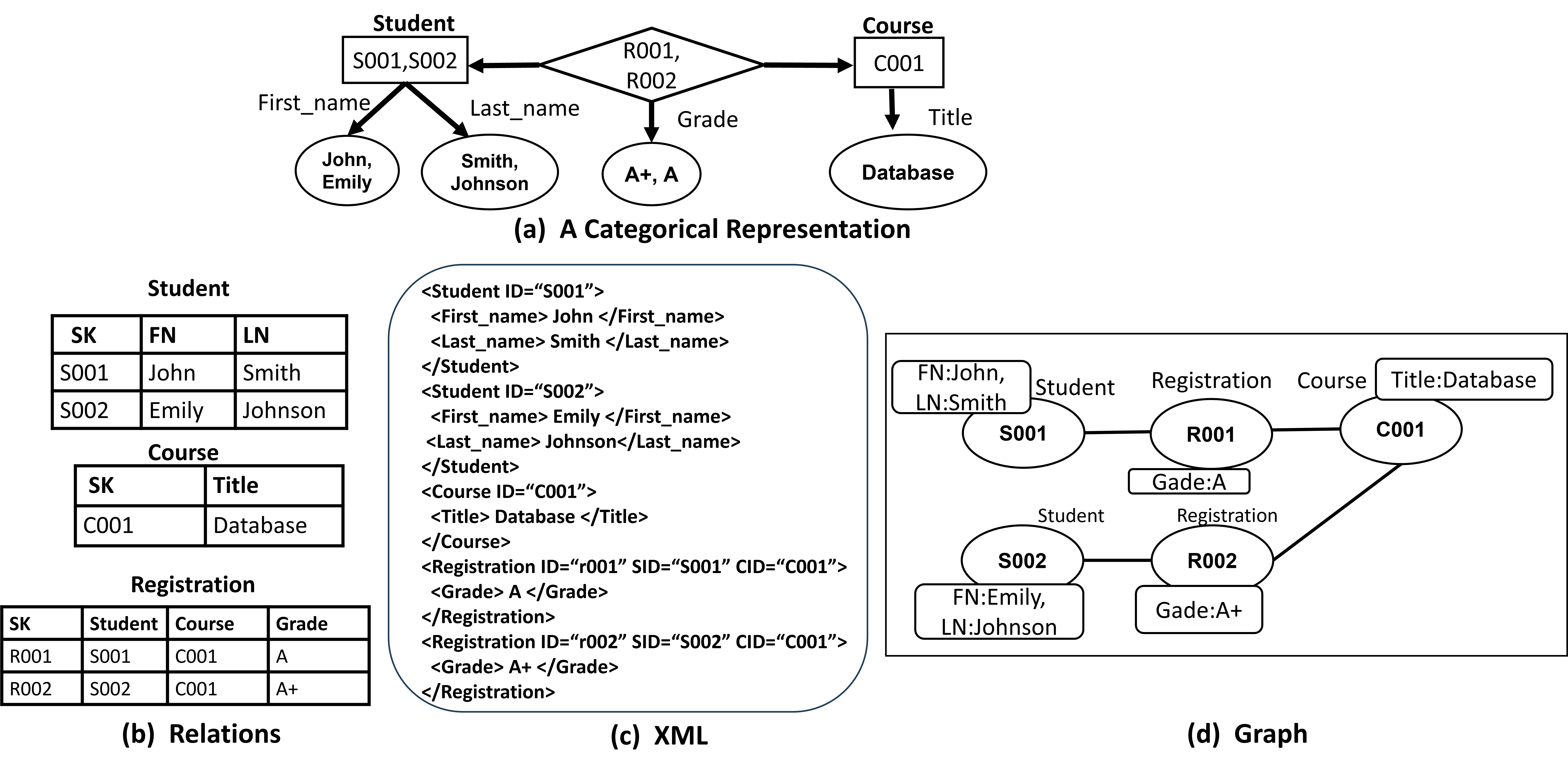}
\caption{The relation, XML and graph data converted from the category in Figure \ref{fig:firstexamplecategory}.} \label{fig:relationXML}
\end{figure*}

\section{Mapping Categories to Graphs}
\label{sec:PropertyGraph}

\begin{algorithm} \caption{Map to Property Graph Schema}\label{alg:map2graphschema}\KwIn{A graph representation $\mathcal{G(C)}$ of schema of a category $\mathcal{C}$ }
    \KwOut{A property graph  schema $\mathcal{PG}$= ($V, E, T, P$)}
    \DontPrintSemicolon

Initialize $V, E, T$, and $P$ to be empty sets;

\ForEach{ object $O$ in $V(\mathcal{G(C)})$ with outgoing edges}
{
     $V \leftarrow V \cup \{\lambda(O)\}$;

     \If{$O$ is an entity or attribute object  }
     {
        $P(O) \leftarrow P(O) \cup \{SK\}$;
     }
    
     Add\_neighbours($O$);
     



    Mark $O$ as a processed object; 

}

\SetKwFunction{func}{ $\mbox{Add\_neighbours}$}
\SetKwProg{Fn}{Function}{:}{}
\Fn{\func{$O$}}
{
    \ForEach{ object $N$ in $outNbr(O)$ }
    {
        \If{$N$ is an attribute object and $N$ has no outgoing edges} 
        { 
            $P(O) \leftarrow P(O) \cup \{\lambda(N)\}$; 

            $T \leftarrow T \cup \{\lambda(N)\}$; 
        }
        \Else
        { 
            $E \leftarrow E \cup \{(O,N)\}$;

            
      
        }

        



    }
    
}   

        





            

                

\end{algorithm}



\begin{definition} 
A (undirected) property graph schema is a 4-tuple ($V, E, T, P$) where 

\begin{itemize}[noitemsep,topsep=0pt]
  \item $V$ is a ﬁnite set of labels of vertices;
  \item $E \subseteq V \times V$, a finite set of labels of edges; 
  \item $T$ is a ﬁnite set of attributes;
  \item $P$ is a function, which maps a label in $V \cup E$ into a subset of $T$.
\end{itemize}
  
\end{definition}



 

Algorithm \ref{alg:map2graphschema} provides a mapping procedure designed to convert a categorical schema into a property graph schema.  The algorithm operates as follows: First, each object with outgoing edges in the category is transformed into a node within the property graph (Lines 2-7). In the case where the object represents an entity or a relationship, its surrogate key is added as a property, denoted by $SK$. Next, after any object $O$ is processed,  if its neighbor object $N$  is an attribute object without outgoing edges, it is added as an attribute for $O$ in the graph (Lines 10-12), otherwise, a new edge is created to connect $N$ and $O$ in the graph (Lines 14).

\begin{example}
Given the graph representation shown in Figure \ref{fig:1RRExample}(c), we use the above algorithm to convert it to a graph schema as follows: nodes $V= \{A,B,D\}$, edges $E= \{(A,B),(D,A)\}$, properties $T=\{SK, C, E\}$, $P(B)=\{SK, C\}$, $P(D)=\{SK, E\}$, and $P(A)=\{SK\}$.
\end{example}

\begin{example}
 Consider the graph representation of Figure \ref{fig:2RRExample}(c), we apply Algorithm \ref{alg:map2graphschema} again: nodes $V= \{A,B,X_1,X_2\}$, edges $E= \{(X_1,A),(X_1,B),(X_2,A)\}$, attributes $T=\{SK, C,D\}$, $P(A)=\{SK,C\}$,  $P(B)=\{SK,C\}$, $P(X_2)=\{SK,D\}$.
\end{example}

In addition, Figure \ref{fig:relationXML} shows another example for the output relations, XML and graph data based on the category from Figure \ref{fig:firstexamplecategory}.




\section{Mapping to Hybrid Schema}
\label{sec:hybrid}

\begin{figure}
\centering
\includegraphics[width=0.7\textwidth]{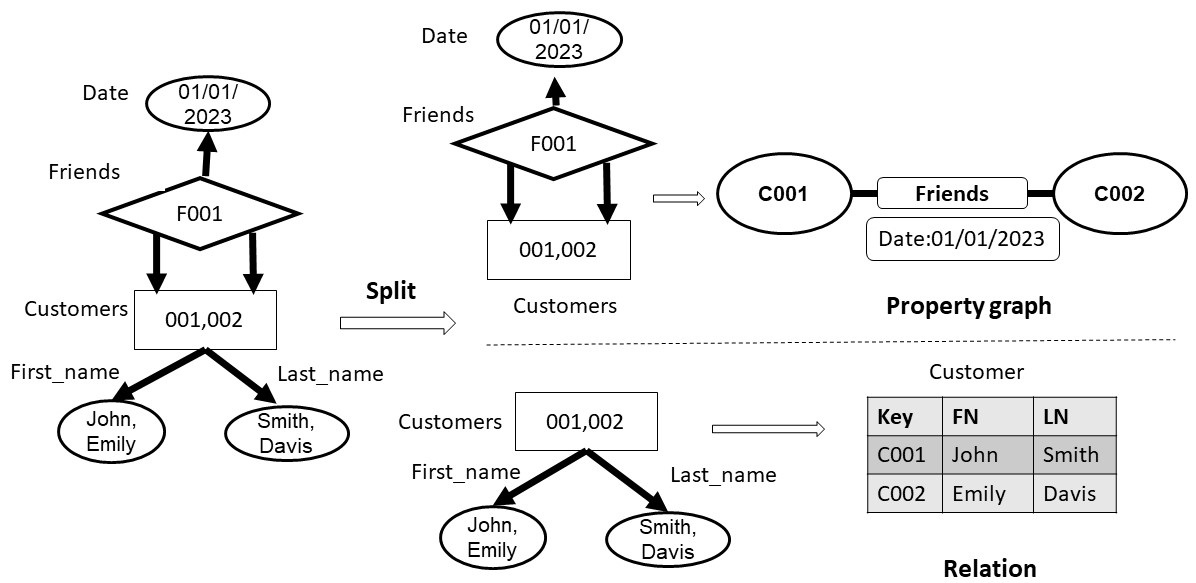}
\caption{An example to illustrate the output of two different models of data} \label{fig:hybrid}
\end{figure}


 In the context of a multi-model database, it is often necessary to decompose a categorical schema into multiple components,  each of which may possess its own unique structure and model, serving specific aspects of the data. To illustrate this concept, Figure \ref{fig:hybrid} demonstrates an example wherein customer data is divided into two distinct parts. The first part pertains to the information of customers, which is represented using the relational data model. The second part captures the friendship connections between customers and is best represented as a property graph model.

 Given a reduced representation $G$ of a category, we aim to decompose this representation $G$ into a set of subgraphs denoted as ${G_1, G_2, ..., G_n}$. The objective is to ensure that each node and edge in $G$ appear in at least one subgraph $G_i$, thereby preserving all information. Furthermore, when an edge $e$ belongs to a category $G_i$, the two nodes connected by $e$ must also be contained within $G_i$. This lossless decomposition approach guarantees that all nodes and edges can be retained after performing the decomposition.

 

\section{Inference rules for FDs and MVDs}
\label{sec:monoidal}

\subsection{Inference Rules for FDs and Monoidal Category}


In Section \ref{sec:ClosureFD}, we introduced three inference rules governing functional dependencies within the framework of Armstrong's axioms. More specifically, the first rule establishes the concept of a projection morphism, while the third rule pertains to a composed morphism. In this appendix, we aim to demonstrate that the second rule in the following, in fact, delineates the properties of a monoidal product.

FD 2: If $f: X \to Y$, then $g: XZ \to YZ$;  Specifically, given an element $(x,z) \in XZ$, let $y = f(x)$, we define that $(y,z) \in YZ$ and $(y,z)$ is the image of $(x,z)$ under the function $g$.

\begin{definition}(Monoidal Category) A monoidal category $\mathcal{C}$ consists of the following components:

\begin{itemize}
    \item A category $\mathcal{C}$ with objects denoted as $X, Y, Z, \ldots$ and morphisms between objects.
    \item A bifunctor $\otimes: \mathcal{C} \times \mathcal{C} \to \mathcal{C}$, called the monoidal product, which associates to each pair of objects $X$ and $Y$ an object $X \otimes Y$ in $\mathcal{C}$.
    \item An associator, which is a natural isomorphism:
    $ \alpha_{X, Y, Z}: (X \otimes Y) \otimes Z \xrightarrow{\sim} X \otimes (Y \otimes Z)$
    satisfying certain coherence conditions.
    \item A unit object $I$ and natural isomorphisms, called the left and right unitors: $ I \otimes X \xrightarrow{\sim} X$ and $  X \otimes I \xrightarrow{\sim} X$
    also satisfying coherence conditions.
\end{itemize}
\end{definition}


Let us consider two distinct categories: $\mathcal{C}_1$, which consists of two objects, denoted as $X$ and $Y$, connected by a morphism $f: X \to Y$, and $\mathcal{C}_2$, featuring a single object $Z$ with an identity morphism. We aim to construct a monoidal category by combining $\mathcal{C}_1$ and $\mathcal{C}_2$, yielding $\mathcal{C}_3$. To achieve this, we introduce a bifunctor denoted as $\otimes: \mathcal{C}_1 \times \mathcal{C}_2 \to \mathcal{C}_3$. This bifunctor associates the pair $(X,Z)$ with the object $XZ$, and the pair $(Y,Z)$ with the object $YZ$. The morphism between objects $XZ$ and $YZ$ is defined as $k: \forall (x,z)\in XZ \to (y,z) \in YZ$, where $f(x)=y$.  Further,  a unit object $I$ is represented as an empty object denoted by $\epsilon$, accompanied by an identity morphism. This unit object $\epsilon$ satisfies the condition that for all elements $x$ within the object $X$, both pairs $(x,\epsilon)$ and $(\epsilon,x)$ correspond to $x$.






\subsection{Inference Rules for MVDs} \label{sec:MVDInference}

Given a set of functional dependencies $F$ and a set of multivalued dependencies $M$, the inference rules to compute their closure can be found in literature, see e.g. \cite{10.5555/551350,10.1145/320613.320614,10.1145/509404.509414}. We provide those inference rules as follows. 
  
MVD 1: (Complement) If $X \to\to_U Y$, then $X \to\to_U (U-XY)$;

MVD 2: (Reflexivity) If $Y \subseteq X$ in a relation $U$, then $X \to\to_U Y$;

MVD 3: (Augmentation) If $Z \subseteq W$ and $X \to\to_U Y$, then $XW \to\to_U YZ$, where $U = X \cup Y \cup Z \cup W$;

MVD 4: (Transitivity) If $X \to\to_U Y$ and $Y \to\to_U Z$, then $X \to\to_U Z-Y$, where $U = X \cup Y \cup Z$.

There are two additional rules for both FD and MVD.

FD-MVD 1: If $X \to Y$ in a relation $U$, then $X \to\to_U Y$.

FD-MVD 2: If $X \to\to_U Z$ and $Y \to Z'$, ($Z' \subseteq Z$), where $U=X\cup Y \cup Z$, $Y$ and $Z$ are disjoint, then $X \to Z'$. 

For FD-MVD 2, given an element $x \in X$, we need to find a unique element $z' \in Z'$.  Without loss of generality, assume that $\exists y \in Y$ s.t.  $(x,y,z)\in U $, then since $Y \to Z'$, we can find this unique element $z' \in Z'$, s.t. $z'$ is the image of $y$ and $x$. Note that it is possible that there are multiple $y$'s associated with $x$ in $U$, but all those $y$'s should map to the same $z'$, otherwise, it contradicts with $X \to Z'$.

The detailed algorithm (by constructing the dependency basis) for computing the membership and closure of FDs and MVDs using the above rules can be found in previous work (e.g., \cite{10.1145/320613.320614,10.1145/322186.322190}).
 

\section{Proof and Explanation of Lemmas and Theorems}
\label{sec:proofs}

\noindent \textit{Lemma \ref{lem:thincommutative}:  All diagrams in a thin category are commutative. }

\begin{proof} (Sketch) Without loss of generality, consider three objects $A$, $B$ and $C$ in a category.  Let $f: A \to B$,  $g: B \to C$, and   let $k = g \circ f$. Since there is only one arrow (i.e. $k$) between $A$ and $C$, then this diagram is commutative, otherwise, there are at least two arrows between $A$ and $C$, which contradicts the definition of a thin category. This commutative result can be proved for any two sequences of objects with the same starting and ending points, which concludes this proof.
\end{proof}

\noindent \textit{Theorem \ref{theo:BCNF}:  The relation schema $R$ output from the first reduced representation in Algorithm  \ref{alg:map2relationschema} is in the Boyce-Codd normal form.}

\begin{proof} 
(Sketch) We aim to prove that every non-trivial functional dependency (FD) in the relation $R$ is a key constraint. To establish this, we will employ a proof by contradiction. Assume the existence of an FD: $A \to B$ in the relation $R$, where $A$ is not a superkey, there are two cases:
(1)  $A$ has a single attribute.  Note that $R$ has at least one single attribute key $K$. This key, by definition, guarantees that $K \to A$ and $K \to B$. If there is an FD $A \to B$, then $K \to B$ should be removed in the minimal cover, which contradicts the fact $B \in R$. Note that the $K$ is possibly removed in \texttt{Clean} Function of Algorithm \ref{alg:map2relationschema}. But this does not change the above proof and the contradiction still occurs. (2) $A$ has multiple attributes, say $A_1 A_2... A_n \to B$. In the algorithm to compute the closure of a category (i.e. Algorithm \ref{alg:closureFD}, Lines 4 and 5), a new object $X$ with the associated projection arrows and  $X \to B$ has been added to the category. Then in the minimal cover, $K \to B$ would have been removed. This also contradicts that $B \in R$.

Furthermore,  note that in the output of Algorithm \ref{alg:map2relationschema}, it is possible for a relation $R$ to contain multiple candidate keys. This is due to the inclusion of bijective objects and their neighbors in $R$ (as indicated in Lines 15-17 of Algorithm \ref{alg:map2relationschema}). Each of these bijective objects corresponds to a candidate key within $R$.
\end{proof}

The inadequacy of BCNF (Boyce-Codd Normal Form) when applied to multiple relations has been demonstrated by Ling et al. \cite{journals/tods/LingTK81}. They identified that BCNF may contain ``\textit{superfluous}''  attributes. In response, the authors proposed an enhanced normal form in their paper, the following example illustrates the limitations of BCNF.

\begin{example} (\cite{journals/tods/LingTK81})
Let F =\{$AB \to CD$, $A \to E$, $B \to F$, $EF \to C$ \}, and consider the relational schema table $T_1(A,B,C,D)$ and key $AB$, $T_2(A,E)$ and key A, $T_3(B,F)$ and key B, and $T_4(E,F,C)$ and the key EF.  Although all tables satisfy the BCNF, a superfluous attribute is $C$ in $T_1$. This example shows the limit of BCNF to identify the redundant attributes across relations. \label{exp:improve3NF} 
\end{example}

\begin{definition} (\cite{journals/tods/LingTK81}) Given a set of relations $S = \{R_1,R_2,...,R_n\}$,  an attribute $B \in R_i$ is restorable in $S$ if $B \notin  K$ and  $K_i \to B$ is a functional dependency that can be derived by other FDs that do not involve $R_i$, where $K$ is a set of keys in $R_i$ and $K_i \in K$.
\end{definition}

In the above definition,  an attribute in a relation schema $R_i$ is \textit{restorable} means that the value of this attribute is derivable from the rest of $S$. In Example \ref{exp:improve3NF}, the attribute C in $T_1$ is restorable (superfluous), as its value is restorable from other tables. That is $AB \to C$ can be derived by $T_2$, $T_3$, and $T_4$, where $A \to E$, $B \to F$, and $EF \to C$.

\begin{definition} Given a set of relations $S = \{R_1,R_2,...,R_n\}$,  a relation schema $R_i \in S$ is in improved Boyce-Codd normal form if (i) $R_i$ is in BCNF; and (ii) no attribute of $R_i$ is restorable in $S$.    
\end{definition}

\noindent \textit{Theorem \ref{theo:improvedBCNF}   The relation schema $R$ output from the first reduced representation in Algorithm  \ref{alg:map2relationschema} is in the improved  Boyce-Codd normal form.}

\begin{proof} (Sketch) The proof of Theorem \ref{theo:BCNF} establishes that the output is guaranteed to be in BCNF. Building upon this result, we proceed to demonstrate that the output relations also do not possess any restorable attributes. In Algorithm \ref{alg:map2relationschema}, an attribute $B \in R_i$ only if there is an arrow from $R_i$ to $B$. In Algorithm \ref{alg:1RR}, note that Line 3 involves the computation of a minimal cover by considering all functional dependencies and arrows within the category. As a result, the minimal cover obtained is globally minimal, ensuring that each attribute $B$ in $R_i$ is not a superfluous attribute.   This concludes the proof.
\end{proof}

Following the exploration of the properties of relational data, we now shift towards examining the normal form for XML data.  We are able to establish proof that the resulting XML data, derived from the first reduced presentation, adheres to XML normal form as defined in the paper by Arenas and Libkin \cite{journals/tods/ArenasL04}.

Let us introduce some concepts about XML normal form from \cite{journals/tods/ArenasL04}.  A tree tuple $t$ in a DTD $D$ is a function that assigns to each path in $D$ a value of an XML tree vertex or a string value. A tree tuple represents a finite tree with
paths from $D$ containing at most one occurrence of each path. For example, consider a DTD with expressions: $\epsilon \to student^+$ and $student \to @ID, BirthYear, Age$. An example of a tree tuple $t$ defines that: 
$t$($\epsilon.student$) = $v_1$, $t$($\epsilon.student.@ID$) = ``1'', $t$($\epsilon.student.BirthYear$.\textsc{\#P}) = ``2000'', etc, where the reserved symbols  \textsc{\#P} represent element type declarations \#PCDATA in DTD.

\begin{figure}
\centering
\includegraphics[width=0.7\textwidth]{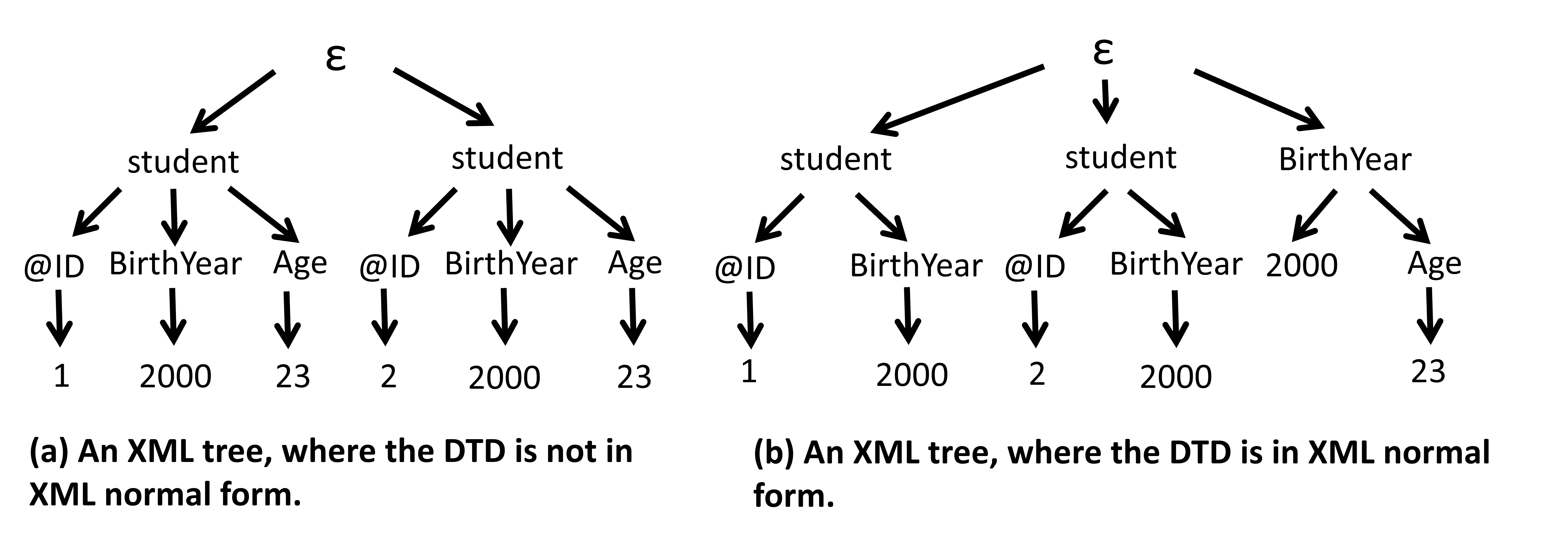}
\caption{An example to illustrate the XML normal form} \label{fig:XMLNFExample}
\end{figure}

For a DTD D, a functional dependency (FD) over $D$ is an expression of the form $S_1 \to S_2$ where $S_1$ and  $S_2$ are nonempty sets of paths in $D$. For example, ``$\epsilon$.student.@ID $\to$ $\epsilon$.student.BirthYear.\#P" is an FD, meaning that each ID of a student can decide the birth year of this student. An XML tree $T$ satisfies $S_1 \to S_2$  if for every two tree tuples $t_1$, $t_2$, $t_1(S_1) = t_2(S_1)$ implies $t_1(S_2) = t_2(S_2)$ in $T$. 
For example,  see an XML tree of Figure \ref{fig:XMLNFExample}. It satisfies  ``$\epsilon$.student.BirthYear.\#P $\to$ $\epsilon$.student.Age.\#P''. But it does not satisfy ``$\epsilon$.student.BirthYear.\#P $\to$ $\epsilon$.student.Age", because the ``BirthYear'' cannot decide a unique ``Age'' vertex. As shown in Figure \ref{fig:XMLNFExample}, there are two ``Age'' vertexes associated with ``BirthYear 2000''. 

Given a DTD $D$, a set $\Sigma \subseteq FD(D)$ and $\varphi \in FD(D)$, we say that ($D, \Sigma$)
implies $\varphi$, written ($D, \Sigma$) $\vdash \varphi$ , if for any XML tree $T$ conforming to $D$ and satisfying $\Sigma$, then $T$ also satisfies $\varphi$. The set of all FDs implied by ($D, \Sigma$) is denoted by ($D, \Sigma$)$^+$.

\begin{definition} (\cite{journals/tods/ArenasL04})
Given a DTD $D$ and $\Sigma  \subseteq$  FD($D$), ($D$, $\Sigma$) satisfies XML normal form if and only if for every nontrivial FD: $\varphi \in$ ($D$, $\Sigma$)$^+$  of the form $X$ $\to$ $p$.@ID or $X$ $\to$ $p$.\#P, they must also have that $X$ $\to$ $p$ is in ($D$, $\Sigma$)$^+$, where $X$ is a finite nonempty subset of paths in $D$, and $p$ is one path in $D$. 
\end{definition}

\begin{example} For example, recall a DTD with expressions: ``$\epsilon \to Student^+$'' and ``$Student \to @ID, BirthYear, Age$'', where XML FDs include that ``$\epsilon$.$Student.@ID$ $\to$ $\epsilon$.$Student.BirthYear.\#P$'',  ``$\epsilon$.$Student.@ID$ $\to$ $\epsilon$.$Student.Age.\#P$'' and ``$\epsilon.Student.BirthYear.\#P$ $\to$ $\epsilon.Student.Age.\#P$''. This DTD is not in an XML normal form. See an example XML tree in Figure \ref{fig:XMLNFExample}(a). Because it is not the case ``$\epsilon.Student.BirthYear.\#P$ $\to$ $\epsilon.Student.Age$'', as there are two $Student$ vertexes with the same $BirthYear$, but different $Age$ vertexes.  To remedy this issue, the $Age$ tag is removed from the children of the $Student$. Consider another DTD: ``$\epsilon \to Student^+ BirthYear^+$'', ``$Student \to @ID, BirthYear$'' and ``$BirthYear \to Age$''. This is in XML normal form, which avoids redundant storage. The corresponding XML  tree is shown in Figure \ref{fig:XMLNFExample}(b).
\end{example}

\begin{lemma}
Given an arrow $f: O_1 \to O_2$ in a representation of a category $\mathcal{C}$, by using Algorithm  \ref{alg:map2DTD} to convert $\mathcal{C}$ to a DTD schema $D$, the corresponding XML FD in $D$ is of the form $p.O_1 \to p.O_1.O_2$  where $p$ is a  path in $D$.
\label{lem:NFPrepare}\end{lemma}

\begin{proof} (Sketch)
We will consider two cases regarding the structural relationship between $O_1$ and $O_2$ in an XML tree that conforms to the output DTD $D$, following the conversion by Algorithm \ref{alg:map2DTD}.

Case (1): $O_1$ is not the parent of $O_2$ in the output XML tree. In this scenario, we have $p.O_1 \to p.O_2$, where both $O_1$ and $O_2$ share the same prefix path $p$, and they are sibling nodes in the XML tree. To demonstrate that this case is not possible, we will employ a proof by contradiction. Let us assume that $O_3$ is a node that represents the common parent of $O_1$ and $O_2$ in the tree. Consequently, we have $O_3 \to O_1$ and $O_3 \to O_2$ in the category. However, this contradicts the fact that the first reduced representation is a minimal cover of arrows, as the existence of $O_3 \to O_1$ and $O_1 \to O_2$ would imply $O_3 \to O_2$. Hence, this case is not possible.

Case (2): $O_1$ is the parent of $O_2$ in the XML tree.
In this situation, the output functional dependency  is of the form: $p.O_1 \to p.O_1.O_2$, which is the desired outcome of the mapping process.

\end{proof}

\noindent \textit{Theorem \ref{theo:DTDNormalform}. The DTD schema output D from the first reduced representation in Algorithm \ref{alg:map2DTD}  is in XML normal form.}

\begin{proof}
(Sketch) We aim to prove that $X \to p.\#P$ or $X \to p.@ID$ implies $X \to p$ in the output DTD.
Let us assume that the DTD schema $D$ is derived from a representation $G$ of a category using Algorithm \ref{alg:map2DTD}. We denote the last two objects in the paths of $X$ and $p$ as $O_1$ and $O_2$ in $G$, respectively. Since $G$ contains the arrow $O_1 \to O_2$, we can infer the existence of the corresponding XML functional dependency $p'.O_1 \to p'.O_1.O_2 \in (D, \Sigma)$ based on Lemma \ref{lem:NFPrepare}.

Now consider the following four cases about the path structure of $X$:

Case (1) $X= p'.O_1$: In this case, $X \to p'.O_1.O_2$. Note that  $p'.O_1.O_2 \to p'.O_1.O_2.@ID$ and  $p'.O_1.O_2 \to p'.O_1.O_2.\#P$. Thus, $X \to p$ belongs to the closure of ($D$, $\Sigma$)$^+$.

Case (2) $X= p'.O_1.@ID$: We have $p'.O_1.@ID \to p'.O_1$, as @ID is a global unique attribute for each element in $O_1$. By the composed rule of arrows,  ($X \to p) \in$ ($D$, $\Sigma$)$^+$.

Case (3) $X= p'.O_1.\#P$: Similarly, we have $ p'.O_1.\#P \to p'.O_1$, as $O_1$ is an attribute object and  different elements in $O_1$ have distinct \#$P$ values. Meanwhile, each $O_1$ is processed only once in Algorithm \ref{alg:map2DTD}. Thus, ($X \to p) \in$ ($D$, $\Sigma$)$^+$.

Case (4) $X= p'.O_1.@l$: where $@l$ is an attribute under $O_1$ except @ID.  We prove that this case is not valid. Without loss of generality, assume this $@l$ attribute represent an attribute object $O_3$ in $G$.  That is, $X= p'.O_1.@O_3$. In other words, $O_1 \to O_3$. Further, note that $O_3 \to O_2$, as $X \to p.@ID$ or $X \to p.\#P$ and $O_2 \in p$. This contradicts that $O_1 \to O_2$ is in 1RR, as it is a redundant arrow due to the existence of $O_1 \to O_3$ and $O_3 \to O_2$.   Thus, 
$ (p'.O_1.@l \to p'.O_1.O_2.@ID) \notin (D,\Sigma)^+$ and $ (p'.O_1.@l \to p'.O_1.O_2.\#P) \notin (D,\Sigma)^+$.

Hence, in all cases above,  we have demonstrated that $X \to p.\#P$ or $X \to p.@ID$ leads to $X \to p$ in the closure of ($D$, $\Sigma$), which concludes the proof.
\end{proof}


\begin{example} This example illustrates that 1RR results in the schema in XML normal form. Recall Figure \ref{fig:1RRExample}(a).  If we use Algorithm \ref{alg:map2DTD} to convert it to XML DTD, there is an XML FD: $\epsilon.A.B.\#P \to \epsilon.A.C.\#P$. But in this case $\epsilon.A.B.\#P \to \epsilon.A.C$ does not hold, because it is possible that there are two tree tuples $t_1$ and $t_2$, s.t. $t_1(\epsilon.A.B.\#P) = t_2(\epsilon.A.B.\#P)$, but $t_1(\epsilon.A.C) \neq t_2(\epsilon.A.C)$. This output schema is not in XML normal form. However, if we convert Fig.  \ref{fig:1RRExample}(a) to (c) with 1RR, $C$ is not a child of $A$ and the XML FD becomes  $\epsilon.B.\#P \to \epsilon.B.C.\#P$. Note that $\epsilon.B.\#P \to \epsilon.B.C$ holds, as each value of $B$ is stored once as a child of the root. Therefore, the output schema with 1RR satisfies XML normal form. 
\end{example}

Because of the absence of a well-established normal form theory specific to graph data (as far as our knowledge), there exists no analogous property that can be proven for the output of a property graph. Nevertheless, as the framework remains unified, we anticipate that the resulting graph data schemata exhibit favorable characteristics that help mitigate a certain level of redundancy, as illustrated in the following example.

\begin{figure}
\centering
\includegraphics[width=0.6\textwidth]{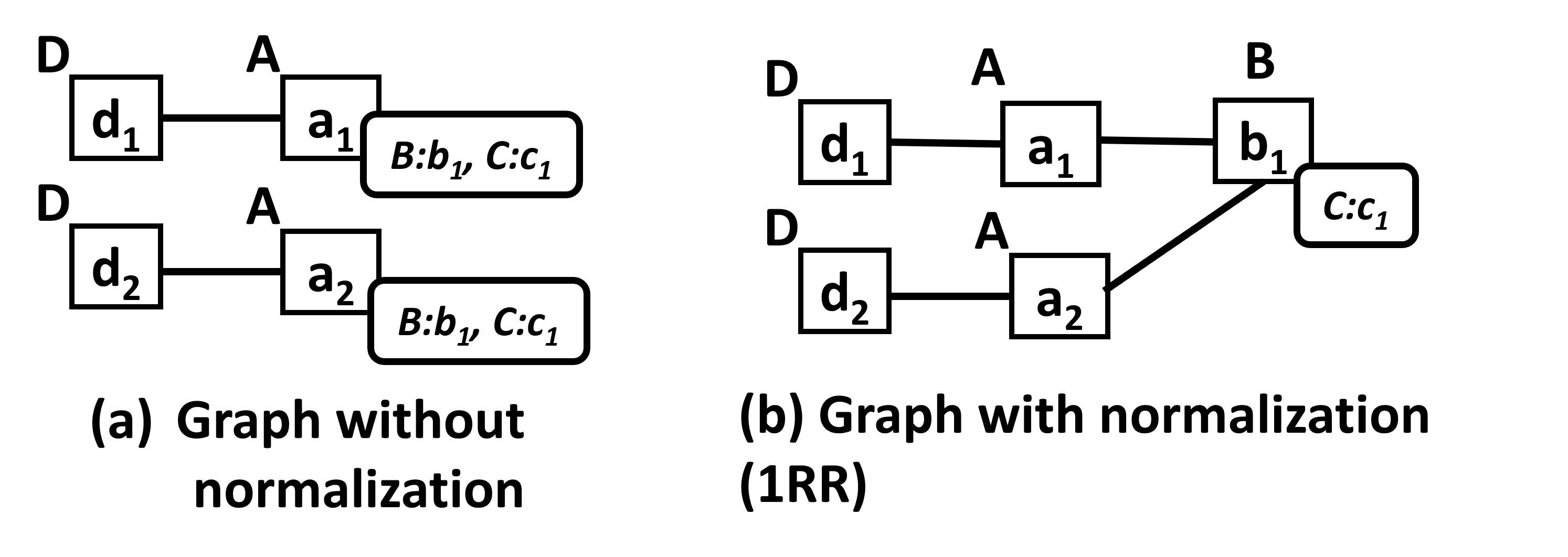}
\caption{An example to illustrate the graph normal form (1RR)} \label{fig:GraphExample}
\end{figure}

\begin{example}
Recall Figure \ref{fig:1RRExample}(a). When applying Algorithm \ref{alg:map2graphschema} to convert this category to a graph schema, we obtain the following schema: nodes $V= \{A,D\}$, edges $E= \{(D,A)\}$, properties $T=\{SK,B,C,E\}$, $P(D)=\{SK, E\}$ and $P(A)=\{SK, B, C\}$. In Figure \ref{fig:GraphExample}(a), we present an example of a graph data instance that adheres to this schema. Note that the properties of $E$ are omitted for simplicity in the figure. In this instance, the values of properties $B$ and $C$ are duplicated and stored for different nodes of type A. However, by adopting the 1RR approach in Figure \ref{fig:1RRExample}(c), we can avoid such redundancy. A graph instance based on the schema from 1RR is shown in Figure \ref{fig:GraphExample}(b), where a new node with type $B$ is created, and $b_1$ and $c_1$ occur only once. This example intuitively demonstrates the benefits of using 1RR in graph schema design, effectively reducing redundancy.
\end{example}



\noindent \textit{Theorem \ref{the:4NF}: Each relation schema output from the second reduced representation in Algorithm \ref{alg:map2relationschema} is in the  fourth normal form.}

\begin{proof} Let us first recall the definition of 4NF in the textbook \cite{elmasri2000fundamentals}.

\textbf{Definition}: A relation schema $R$ is in 4NF with respect to a set of dependencies $F$ (that includes functional dependencies and multivalued dependencies) if, for every nontrivial multivalued dependency $X \to \to Y$ in $F^+$, $X$ is a superkey for $R$.

Note that if $X \to Y$ in a relation $U$, then $X \to \to_U Y$. Thus any functional dependency can be also considered as a multivalued dependency. A multivalued dependency $X \to\to Y$ in $R$ is  trivial  if (a) $Y$ is a subset of $X$ or (b) $X \cup Y = R$.   By Theorem \ref{theo:BCNF}, we have already established that the  schema from 1RR follows BCNF.  That is for every functional dependency $X \to Y$ in $F^+$, $X$ is a superkey of $R$. Next we consider the case where the nontrivial multivalued dependencies that are not functional dependencies. In Line 2 of Algorithm \ref{alg:2RR} that computes the second reduced representation, we observe that all multivalued dependencies (MVDs) objects are decomposed.  This implies that there are no such nontrivial multivalued dependencies  $X \to \to_U Y$ in $F^+$ which can be output in a single relation $U$ in Algorithm \ref{alg:map2relationschema}. This concludes the proof.
\end{proof}

\section{Complexity analysis of algorithms}
\label{sec:complexity}

The time complexity of Algorithm \ref{alg:closureFD} is dominated by the cost to compute the closure of the functional dependencies (i.e. Line 4). Let $m$ and $n$ be the number of objects and arrows in $G$, and let $d$ be the number of FDs. The total time for computing the FD closure is $O((d+n) \cdot m)$ based on the implementations of the previous work (e.g. \cite{10.1145/320493.320489}). Using this procedure, one can implement Algorithm \ref{alg:closureFD} with the time $O((d+n) \cdot m)$.

The time complexity of Algorithm \ref{alg:closureMVD} is essentially that of the computation of the closure of FDs and MVDs. Let $m$ be the number of objects in $G$, $n$ the number of arrows in $G$, and let $d_1$ and $d_2$ be the number of FDs and MVDs, respectively. The total time for computing the FD closure is $O((d_1+n) \cdot m)$ and MVD closure is $O((d_2+n) \cdot m^3)$ based on the implementations of papers  (\cite{10.1145/320493.320489, 10.1145/320613.320614}. Using these procedures, one can implement Algorithm \ref{alg:2RR} with the time $O((d_1+d_2+n) \cdot m^3)$.

The time complexities of Algorithms \ref{alg:1RR} and \ref{alg:2RR} are essentially equivalent to those of computing the closures of the categories. In other words, they mirror the time complexities of Algorithm \ref{alg:closureFD} and \ref{alg:closureMVD}, respectively, which have been previously analyzed.

Algorithms \ref{alg:map2relationschema}, \ref{alg:map2DTD} and \ref{alg:map2graphschema} demonstrate a computational complexity of $O(m+n)$, where $m$ denotes the number of objects in $G$, $n$ the number of arrows in $G$, as each arrow is processed once.

\end{document}